\documentclass[paper]{aa}

\usepackage{graphicx}
\def\ls{{_<\atop^{\sim}}}

\def\cgs{ ${\rm erg~cm}^{-2}~{\rm s}^{-1}$ } 

\begin{document}
\title{The XMM-Newton survey of the ELAIS-S1 field. I: Number counts,
angular correlation function and X-ray spectral
properties.\thanks{based on observations made with XMM-Newton, an ESA
science mission.}}

\author{S. Puccetti\inst{1,2}, F. Fiore\inst{1}, V. D'Elia\inst{1}, I. 
Pillitteri\inst{3}, C. Feruglio\inst{1}, A. Grazian\inst{1}, M. Brusa\inst{4}, 
P. Ciliegi\inst{5}, A. Comastri\inst{5}, 
C. Gruppioni\inst{5},  M. Mignoli\inst{5}, C. Vignali\inst{6}, 
G. Zamorani\inst{5}, F. La Franca\inst{7}, N. Sacchi\inst{7},  
A. Franceschini\inst{8}, S. Berta\inst{8}, 
H. Buttery\inst{9}, J.E. Dias\inst{9}}

\institute {INAF-Osservatorio Astronomico di Roma \\
via Frascati 33, Monteporzio-Catone (RM), I00040 Italy.
\email{puccetti@mporzio.astro.it}
\and
Dip. di Fisica Universit\`a di Roma Tor Vergata
\and
Dip. di Fisica Universit\'a degli Studi di Palermo
\and
Max Planck Institut f\"ur Extraterrestrische Physik (MPE), Giessenbachstrasse 1, 
D-85748 Garching bei M\"unchen, Germany  
\and
INAF-Osservatorio Astronomico di Bologna
\and
Dip. di Astronomia, Universit\`a di Bologna
\and
Dip. di Fisica, Universit\`a Roma Tre
\and
Dip. di Astronomia, Universit\`a di Padova
\and
INAF-Osservatorio Astrofisico di Arcetri
}

\date{July 6 2006}

\abstract
{}
{The formation and evolution of cosmic structures can be probed by
studying the evolution of the luminosity function of AGNs, galaxies and
clusters of galaxies and of the clustering of the X-ray active
Universe, compared to the IR-UV active Universe.}
{To this purpose, we have surveyed with XMM-Newton the central
$\sim0.6$ deg$^2$ region of the ELAIS-S1 field down to flux limits of
$\sim 5.5\times 10^{-16}$ \cgs (0.5-2 keV, soft band, S), $\sim
2\times 10^{-15}$ \cgs (2-10 keV, hard band, H), and $\sim 4\times
10^{-15}$ \cgs (5-10 keV,ultra hard band, HH).}
{We detect a total of 478 sources, 395 and 205 of which detected in
the S and H bands respectively.  We identified 7 clearly extended
sources and estimated their redshift through X-ray spectral fits with
thermal models. In four cases the redshift is consistent with z=0.4. We
have computed the angular correlation function of the sources in the S
and H bands finding best fit correlation angles $\theta_0=5.2\pm3.8$
arcsec and $\theta_0=12.8\pm7.8$ arcsec, 
respectively. A rough estimate of the present-day
correlation length r$_0$ can be obtained inverting the Limber equation and 
assuming an appropriate redshift distribution dN/dz. The results 
range between 12.8 and 9.8 h$^{-1}$ Mpc in the S band and between
17.9 and 13.4   h$^{-1}$ Mpc in the H band, with 30-40\% statistical errors,
assuming either smooth redshift distributions or redshift distributions with 
spikes accounting for the presence of significant structure at z=0.4.
The relative density of the S band sources is higher near the clusters
and groups at z$\sim0.4$ and extends toward East and toward
South/West. This suggests that the structure is complex, with a size
comparable to the full XMM-Newton field. Conversely, the highest
relative source densities of the H band sources are located in the
central-west  region of the field.
}
{}

\authorrunning {Puccetti et al.}
\titlerunning {The XMM-Newton survey of the ELAIS-S1 field}

\maketitle

\section{Introduction}

One of the most challenging goals of modern cosmology is to understand
how the structures in the Universe have been formed and evolved during the
time. Both clusters of galaxies and Active Galactic Nuclei (AGNs) shine
powerfully in the X-ray band and therefore X-ray surveys are the most
efficient tool to both trace the evolution of the cosmic web from low
to high redshift and investigate the correlation between the formation
and light-up of supermassive black holes (SMBH) in AGNs and the
formation and evolution of galaxies. X-ray surveys are less biased
against obscured AGN than optical UV surveys and thus reach higher
surface densities ($\sim1000$ deg$^{-2}$ at the flux limits reachable
by Chandra and XMM-Newton in 50-100 ksec) than even faint optical AGN
surveys (e.g., the COMBO 17 survey reaches $\sim200$ deg$^{-2}$ with
magnitude limit R$\ls24$, Wolf et al. 2004), and so they can be used
to trace the Large Scale Structure (LSS) more efficiently than optical
AGNs.  While there is today no doubt that tight links and feedbacks do
exist between SMBH activity and galaxy evolution and between galaxy
activity (nuclear and star-formation) and the larger scale
environment, very little is known about how these links and feedbacks
work in detail.

One of the most important achievements of Chandra and XMM-Newton
surveys concerns the redshift distribution of the hard X-ray selected
sources. Indeed it has been realized that the accretion powered
luminosity density is dominated by obscured, low luminosity Seyfert
like AGN at redshifts 0.5-1. On average, the activity of Seyfert like
objects rises up to z$\approx1$ and then decreases (Hasinger 2003,
Fiore et al. 2003, Ueda et al. 2003, Barger et al 2005,
\cite{lafranca}), recalling the evolution of star-forming galaxies. On the
other hand, QSO activity rises steeply up to z=2-3 or even further,
recalling the evolution of massive spheroids (Franceschini et al.
1999). The z=0.5--2 redshift range can then be considered as the
``golden epoch of galaxy and AGN activity''.  Both activities are
likely to be due to a) the availability of large masses of cold dust
and gas in galaxies; b) frequent and efficient interactions which
destabilize the gas and make it available for both star-formation and
nuclear accretion (Cavaliere \& Vittorini 2000; Menci et al. 2004).

Large-area, high-sensitivity X-ray surveys can provide a test of the evolution
of cosmic structures along the past light-cone and, therefore, are powerful
probes of the baryonic content of the Universe. The tools to address this
issue are the study of the evolution of the luminosity function of AGN,
galaxies and clusters of galaxies and the study of the clustering of the X-ray
active Universe, compared to the IR-UV active Universe. This comparison will
allow a ``direct'' investigation of how different kinds of galaxy activity
trace the cosmic web. The comparison between AGN clustering, galaxy clustering
and theoretical models for the evolution of the LSS can allow us to understand
whether AGNs trace higher density peaks or higher mass haloes and how active
sources are correlated with the environment.  Deep multiwavelength surveys
like GOODS (i.e. the Great Observatories Origins Deep Survey) sample a volume
of the Universe too small to be representative at z=0.5--2, the epoch where
the action is. Hydrodynamic, $\Lambda-$CDM, simulations for the formation and
evolution of the structure in the Universe show that boxes of the order of 50
Mpc side can be fair samples of the Universe and, consequently, only at these
scales all mass overdensities relative to the mean, will be sampled at their
"true" occurrence rates (see, e.g., \cite{springel}). These linear scales
translate to boxes of 2 and 1.5 degrees side at z=0.5 and z=2, respectively
(assuming a "Concordance" cosmology, i.e. $H_0=70$ km s$^{-1}$ Mpc$^{-1}$,
$\Omega_M$=0.3, $\Omega_{\Lambda}=0.7$ ). As a first step into this direction
we have surveyed with XMM-Newton the central $\sim0.6$ deg$^2$ region of the
ELAIS-S1 field, corresponding to a linear scale of $\sim$18 Mpc at z=0.5.  The
ongoing COSMOS (Scoville et al., in preparation) multiwavelength survey will
be the next step to be pursued in the following several years.  The ELAIS-S1
field (Rowan Robison et al. 2004, Oliver et al. 2000) has been already
surveyed in the mid-IR (15 $\mu$m) by ISO (Lari et al. 2001), in the radio
with the Australia Compact Array (1.4 GHz, Gruppioni et al. 1999), in the
optical/NIR by several ESO and Australian telescopes, which provided deep
optical (Berta et al. 2006) and NIR (Dias et al. in preparation, Buttery et
al. in preparation) photometry in the B, V, R, I, z, and K bands. A campaign
of optical spectroscopic observations is ongoing with VIMOS@VLT (La Franca et
al. 2004). This field is also the main field of the Spitzer legacy program
SWIRE and was observed by Spitzer in Dec 2004 down to flux limits of
3.7$\mu$Jy at 3.6$\mu$m and of 0.15 mJy at 24$\mu$m (Lonsdale et al. 2003,
2004).

This paper presents the XMM-Newton observations of the ELAIS-S1 field,
the number counts in different energy bands and the 2D clustering
properties of the X-ray sources.  The paper is organized as follows:
Sect. 2 presents the data analysis and the X-ray source catalog;
Sect. 3 presents the number counts in four energy bands; Sect. 4
discusses briefly the source X-ray properties; Sect. 5 presents an
analysis of the source clustering and finally Sect. 6 discusses our
main findings.

A $H_0=70$ km s$^{-1}$ Mpc$^{-1}$, $\Omega_M$=0.3, $\Omega_{\Lambda}=0.7$
cosmology is adopted throughout the paper.

\section{The X-ray observations}

A mosaic of four partially overlapping deep XMM-Newton pointings covers
a large ($\sim$0.6 deg$^2$) and contiguous area of the ELAIS-S1
region. The pointings are named ELAIS-S1-A (R.A.=8.91912,
Dec.=-43.31344, J2000), ELAIS-S1-B (R.A.=8.92154, Dec.=-43.65575,
J2000), ELAIS-S1-C (R.A.=8.42195, Dec.=-43.30488, J2000) and
ELAIS-S1-D (R.A.=8.42375, Dec.=-43.65327, J2000). The X-ray
observations were performed on May 2003 through July 2003 with the
European Photon Imaging Camera (EPIC: one PN-CCD camera (0.5-10 keV,
\cite{struder}) and two MOS-CCD cameras (MOS1, MOS2, 0.3-10 keV,
\cite{turner})). Table 1 gives a log of the XMM-Newton observations.

\begin{table*}
\begin{center}
\caption{\bf Observation log}
\begin{tabular}{lcccc}
\hline
Instrument &  Exposure (ksec)& Net exposure (ksec)$^I$ & Energy range & 
Count rate$^{II}$ (s$^{-1}$)  \\
\hline
\multicolumn{5}{c}{ELAIS-S1-A OBSID=0101 $^{III}$}\\
\hline
PN &  84    & 45 & 0.5-10 keV     & 4.165$\pm$0.010  \\
MOS1 & 85    & 51   &0.3-10 keV  &    1.348$\pm$0.005     \\
MOS2  & 85    & 52 & 0.3-10 keV   &  1.352$\pm$0.005     \\
\hline
\multicolumn{5}{c}{ELAIS-S1-B OBSID=0801, 0901, 1001, 1601 $^{III}$ }\\
\hline
PN &   92   & 42 & 0.5-10 keV    &   3.794$\pm$0.012      \\
MOS1 & 100    & 52   &0.3-10 keV  &   1.377$\pm$0.008     \\
MOS2  &  100   & 53 & 0.3-10 keV   &   1.366$\pm$0.008     \\
\hline
\multicolumn{5}{c}{ELAIS-S1-C  OBSID=1201, 1301, 1401, 2101 $^{III}$}\\
\hline
PN &  85    &51 & 0.5-10 keV     & 4.091$\pm$0.009        \\
MOS1 &   97  & 60   &0.3-10 keV  &  1.214$\pm$0.005      \\
MOS2  &  97   & 60 & 0.3-10 keV   &  1.183$\pm$0.005      \\
\hline
\multicolumn{5}{c}{ELAIS-S1-D  OBSID=0401 $^{III}$}\\
\hline
PN &    88  &51 & 0.5-10 keV     &  3.921$\pm$0.009      \\
MOS1 &  89   &  57  &0.3-10 keV  &    1.315$\pm$0.005      \\
MOS2  & 89    & 62 & 0.3-10 keV   &   1.356$\pm$0.005     \\
\hline
\end{tabular}
\end{center}
$^I$Net on-axis exposure time after rejection of high background periods (see
Sect. 2.1); $^{II}$ Total count rate from the whole chip after
rejection of high background periods; $^{III}$Observation IDs.

\end{table*}

\subsection{Data reduction}

The data have been processed using the XMM-Newton Science Analysis
Survey (SAS) v.5.4.1. For the fields ELAIS-S1-A, ELAIS-S1-B and
ELAIS-S1-C, we used the event files linearized with a standard
reduction pipeline (Pipeline Processing System, PPS) at the Survey
Science Center (SSC, University of Leicester, UK).  The PPS data of
the ELAIS-S1-D field are not available in the XMM-Newton archive and therefore we
used the raw event files (i.e. the Observation Data Files, ODF), which
have been linearized with the XMM-SAS pipelines, {\sc epchain} and
{\sc emchain} for the PN and MOS cameras respectively.

The fields ELAIS-S1-B and ELAIS-S1-C were both observed at four
different times.  The event files of the first three observations
(OBSID=0801, 0901, 1001) of the ELAIS-S1-B field and the event files of
all four observations of the ELAIS-S1-C were merged together using the
XMM-SAS task {\sc merge}, because the roll angles are similar (within
0.03$^\circ$). The fourth observation (OBSID=1601) of the ELAIS-S1-B
has a roll angle 0.6$^\circ$ from the other three and therefore it was
analyzed independently.

Events spread at most in two contiguous pixels for PN
(i.e. pattern=0-4) and in four contiguous pixels for MOS (i.e,
pattern=0-12) have been selected.  Event files were cleaned from bad
pixels (hot pixels, events out of the field of view, etc.) and the
soft proton flares.  The soft proton flares are due to solar protons
with energies less than a few hundred of keV. The flares can produce a
count rate up to a factor 100 greater than the mean stationary
background count rate. They are variable during an observation and
from observation to observation, and do not have a predictable
spectral shape or spatial distribution on the detector
(\cite{repo}). In order to remove periods of unwanted high background
level, we located the flares by analyzing the light curves of the
count rate at energies higher than 10 keV, where the X-ray sources
contribution is negligible. We rejected the time intervals when the
count rate is higher than 1.2 counts s$^{-1}$ and 0.3 counts s$^{-1}$
for the PN and MOS cameras respectively. These thresholds maximize the
S/N of faint sources. In conclusion we have rejected $\sim40-50\%$ of
the on source time.

\begin{figure*}
\centering
\includegraphics[angle=0,height=14truecm,width=14truecm]{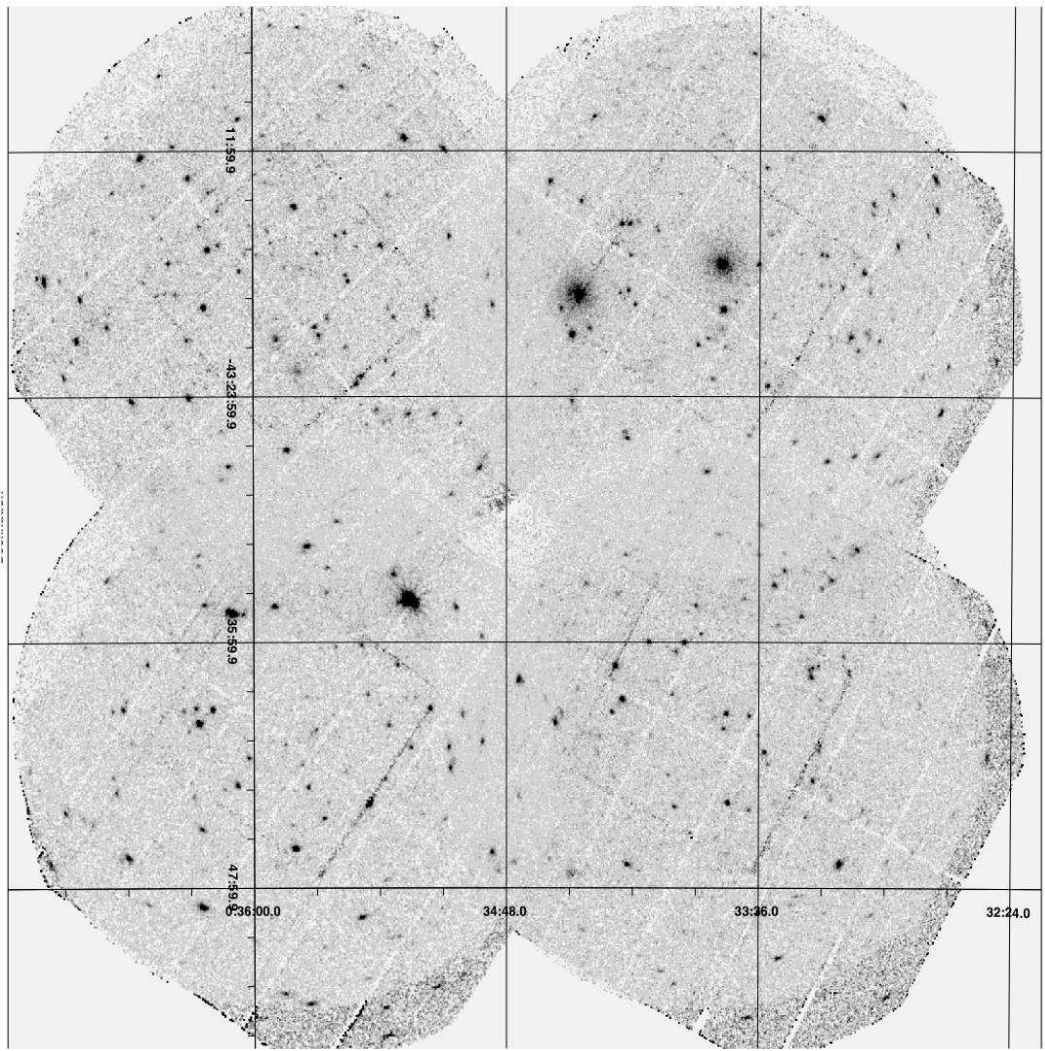}
\caption{Mosaic of the four XMM-Newton pointings (ELAIS-S1-A,
ELAIS-S1-B, ELAIS-S1-C, ELAIS-S1-D) obtained by adding up the images of the
PN and MOS cameras in the energy range 0.5-10 keV. The image is
exposure map corrected. A smoothing factor
with $\sigma=1.5$ pixel (or 6'') was applied. Two out of the three
brightest sources in the field are
extended sources, which are presented in Section 2.5; the other
one is a type 1 AGN at z$=$0.143 (La Franca et al. 2004).}
\label{mosaic}
\end{figure*}

\begin{figure*}
\centering
\includegraphics[angle=0,height=14truecm,width=14truecm]{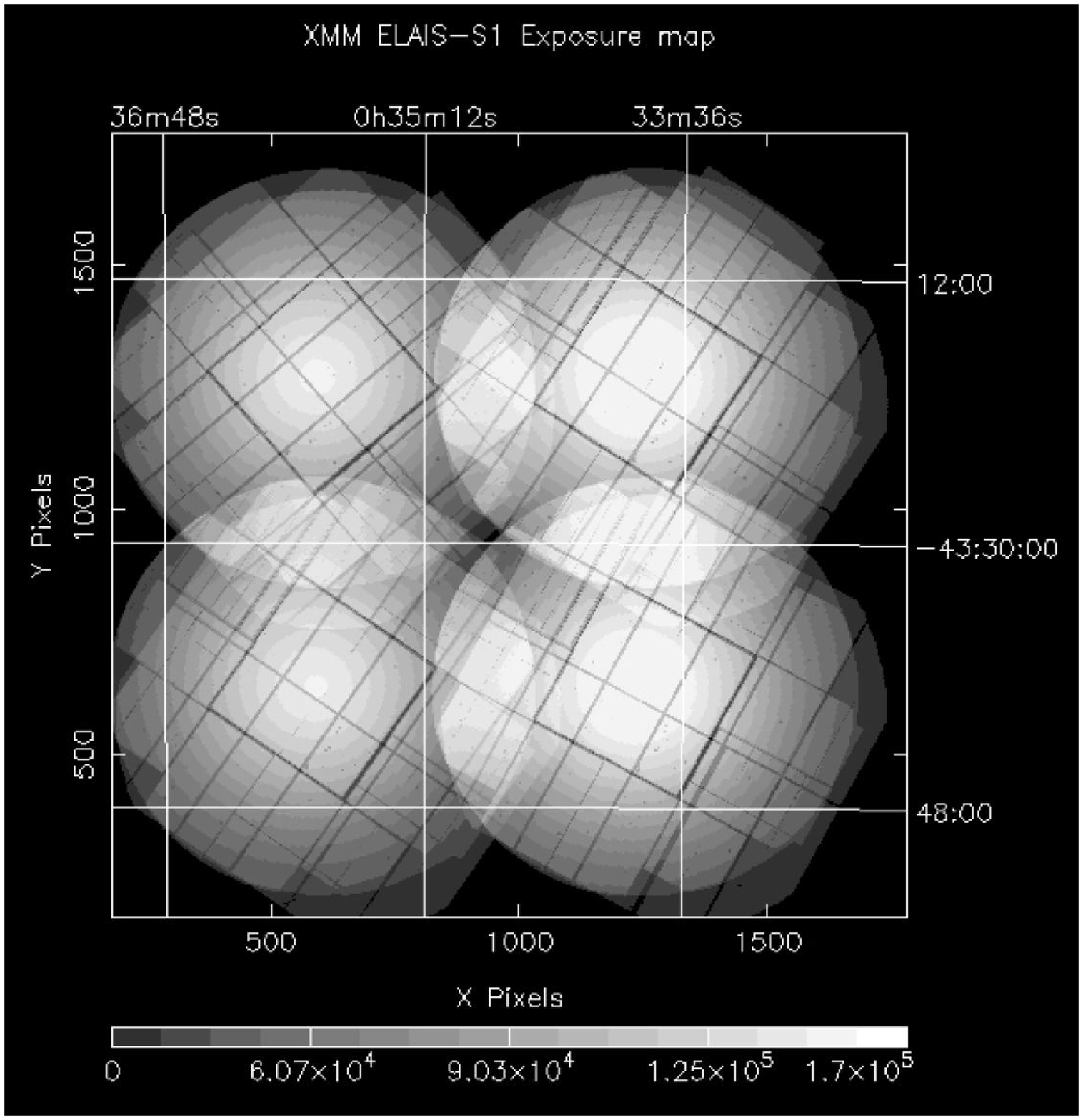}
\caption{0.5-10 keV exposure map of the mosaic of the four XMM-Newton pointings
(ELAIS-S1-A, ELAIS-S1-B, ELAIS-S1-C, ELAIS-S1-D), obtained by adding to the PN
exposure map, the MOS exposure maps weighted on the relative effective area.}
\label{expo}
\end{figure*}

The observations of the four XMM-Newton fields were brought to a
common reference frame by matching the positions of X-ray sources with
bright, point like, optical counterparts with the optical position
given by \cite{lafranca04}. The systematic shifts between the X-ray
and the optical positions were of $\sim$ 2''-4''.  Figure \ref{mosaic}
shows a mosaic of the four XMM-Newton pointings (ELAIS-S1-A,
ELAIS-S1-B, ELAIS-S1-C, ELAIS-S1-D) made adding up the images of the
PN and MOS cameras in the energy range 0.5-10 keV. Figure \ref{expo}
shows the relative exposure map (see Table 1).

\subsection{Source detection}

Source detection was performed on coadded (in sky coordinates)
PN+MOS1+MOS2 images accumulated in four energy bands: 0.5-10 keV (full
band, F), 0.5-2 keV (soft band, S), 2-10 keV (hard band, H), 5-10 keV
(ultra hard band, HH). These bands have been chosen to ease the
comparison with previous work.  The EPIC effective area strongly
decreases after 4-5 keV and therefore the 2-10 keV count rate is
dominated by the count rate at the lower end of the band, unless the
spectrum is extremely hard and/or heavily absorbed. At the flux limit
of our survey we expect that only a minority of the 2-10 keV sources
are highly obscured (see Comastri et al. 2001). In fact, the number of
sources detected in the 2-10 keV band is similar (within 10 \%) to the
number of sources detected in the 2-5 keV band.

The source detections and the X-ray photometry were perfomed by using
the PWXDetect code, developed at INAF~-~Osservatorio Astronomico di
Palermo, following Pillitteri et al. (2006). The code is derived from
the original ROSAT code for source detection by \cite{damiani} and
allows one to combine data from different EPIC cameras and data taken in
different observations, in order to achieve the deepest sensitivity.
The code is based on the analysis of the wavelet transform (WT) of the
count rate image. A WT of a two-dimensional image is a convolution of
the image with a ``generating wavelet'' kernel, which depends on
position and length scale.  In the algorithm developed by
\cite{damiani}, the generating wavelet is a ``mexican hat''. The
length scale is a free parameter; therefore this method is
particularly well suited for the cases in which the point-spread
function (PSF) is varying across the image. It also provides robust
detections of extended sources.

As a first step, a preliminary background map is computed by a
suitable smoothing of the image and using also exposure maps to handle
spatial exposure non-uniformities. Then, a local median filter is
applied to recompute the background at several scales, to minimize the
effect of point sources on the background determination. The WT
algorithm is applied at several wavelet scales. Spatial maxima are
selected if their heights are above the expected background at a
chosen significant level. Information about position, shape and count
rate of the sources are retained for the detection scale at which each
source has the maximum significance. The background is then
recomputed, excluding sources found at the first step and WT procedure
is repeated in order to detect the weakest sources. The final source
list with their positions, detection scales and photometry is then
built. We have chosen a significance level corresponding to a
probability of 2$\times$10$^{-5}$ that a maximum is a Poisson
fluctuations of the background. This corresponds to about one spurious
source per field.  To calibrate this probability we performed 150
simulations of each field in each energy range, assuming a background
equal to the total counts in each field (see last column of Table
1). We then applied the WT algorithm to each simulation and evaluated
the significance level for which we have one spurious source per
field.

To evaluate count rates, we have chosen PN as the reference
detector. The scaling factor (r) between PN, MOS1 and MOS2 depends on
the relative instrument efficiency and source spectral shape. We
evalued this factor for each energy band, from the PN, MOS1, MOS2
count rate ratios, predicted for a power law model with an energy
index $\alpha_E=0.8$.  The count rates in the summed image (CR$_{\rm
Sum}$) are equivalent to PN count rates according to the following
formula:
\begin{small}
\[\mathrm{CR_{Sum}}= \mathrm{CR}_\mathrm{PN} + \mathrm{CR}_\mathrm{MOS1}\cdot 
r_\mathrm{MOS1/PN} + \mathrm{CR}_\mathrm{MOS2}\cdot r_\mathrm{MOS2/PN}
\] 
\end{small}
where the scaling factors $r_\mathrm{MOS1/PN}$ and
$r_\mathrm{MOS2/PN}$ are similar.  The exposure times used to
calculate the count rates are obtained from exposure maps, and thus
they take into account all spatial non-uniformities due to vignetting,
chip geometry, CCD gaps.  We have checked that the source count rates
evalued by the PWXDetect, are consistent with the count rates in
single PN and MOS cameras.

448 sources are detected in the F band images with a probability lower
than 2$\times$10$^{-5}$ that they are Poisson background fluctuations,
the weakest sources having at least 10 net counts in each EPIC
cameras (see Table 2 for more details). The signal to noise ratio of the
weakest detections in the coadded PN+MOS1+MOS2 images is $\sim3$ .
395, 205 and 31 sources are detected in the coadded S, H and HH band
images respectively (see Table 3 for more detail).
Seven of the detected sources are clearly extended. We discuss these
sources in Sect. 2.5. The quality of the detections and of the X-ray
centroids were always checked interactively. We further checked the
quality of the WT algorithm detections by comparing the distributions
of the source counts, count rates, background, signal to noise ratio,
and probability to that obtained using the SAS maximum likelihood
method on the XMM-Newton survey of the COSMOS field (Cappelluti et
al., in preparation), obtaining consistent results.

The source catalog, including position and fluxes, will be
electronically available from VizieR. It is available also at the
following site: http://www.mporzio.astro.it/ELAIS-S1. This site gives
access to a multiwavelenth catalog which also includes optical, near
infrared and mid infrared magnitudes and fluxes and optical
spectroscopy. The optical follow-up campaign is currently on going and
therefore the multiwavelength catalog cannot be part of this
publication and will be properly presented in a future paper (Feruglio et
al., in preparation).

\begin{table}
\begin{center}
\caption{\bf Sources detected in each of the XMM-ELAIS-S1 field}
\begin{tabular}{lccc}
\hline
Energy range & PN counts $^{a}$ & MOS counts $^{b}$ &N$^{c}$\\
\hline
\multicolumn{3}{c}{ELAIS-S1-A}\\
\hline
0.5-10 keV    & 10-450 & 10-350 & 152\\                                  
0.5-2 keV   &  15-370 & 20-250& 135\\ 				        
2-10 keV   &   10-110 & 10-110 & 73 \\ 				        
5-10 keV   & 25-30 &20-25 &13\\ 				        
\hline							        
\multicolumn{3}{c}{ELAIS-S1-B}\\			         
\hline							        
0.5-10 keV       &15-5700 & 15-7000& 106\\ 				          
0.5-2 keV   &15-5100&20-5800 &102 \\ 					         
2-10 keV   &10-600 &10-1100 &62 \\ 					         
5-10 keV   &10-100 &10-140&8\\ 					         
\hline							          
\multicolumn{3}{c}{ELAIS-S1-C}\\			        
\hline							          
0.5-10 keV   & 15-590 &20-450  &157 \\ 				         
0.5-2 keV   & 10-1000 &10-830  & 134 \\ 				         
2-10 keV   & 20-190 &15-140 & 69 \\ 				        
5-10 keV   &15-40 &5-20 &11\\ 					        
\hline							        
\multicolumn{3}{c}{ELAIS-S1-D}\\			         
\hline							        
0.5-10 keV       &20-420 &20-300  &126\\ 				         
0.5-2 keV   &  10-350 &15-260 & 112 \\ 				          
2-10 keV   & 15-130 &15-80  &54\\ 					         
5-10 keV   &  10-35 &10-35  &7\\ 					        
\hline							         
\end{tabular}						        			       
\end{center}						          
							        
Minimum-maximum net counts of the detected sources in the PN
camera ($^{a}$), and MOS1$+$MOS2 cameras ($^{b}$). The counts are not
corrected for vignetting and PSF. ($^{c}$) Number of the detected
sources in each of the four energy bands and in each field. Note that
the four XMM-Newton pointings are partially overlapping, therefore the
sum of the sources in this table is larger than the total number of
the detected sources; note also that the brightest source in the ELAIS-S1-B 
field lies close to a gap between two PN CCDs.

\end{table}

\begin{table}
\caption{\bf Detected sources}
\begin{tabular}{l|c}
\hline
X-ray bands & N$^{a}$\\
\hline
S & 395 \\
H & 205 \\
HH & 31 \\
F & 448 \\
\hline 
F+S+H+HH & 27\\
F+H+HH & 1\\
F+S+HH & 3\\
F+S+H & 140\\
F+H &33\\
H only & 4\\
F+S & 199\\
F only & 45\\
S only &26\\
\hline
\end{tabular}

$^{a}$ Number of detected sources.

\end{table}

\subsection{Count rates, fluxes and sky-coverage}

Source count rates in the four bands were extracted and corrected for
the energy dependent vignetting (as calibrated in orbit by
\cite{lumb}) and PSF (using the analytical approximation given by
\cite{ghizzardia} and \cite{ghizzardib}). The PN and MOS 4 keV PSF
half power diameter (HPD) is $\sim$20 and $\sim$16 arcsec,
respectively, at small ($<$2 arcmin) off-axis angles; it increases
only by a few $\%$ at off-axis angles of 10 arcmin. Even at these
large off-axis angles, the core can be safely considered
axis-symmetric.

For a sample of 40 sources in a broad range of brightness and off-axis
angles, we compared the count rates evaluated using the WT algorithm
to the count rates measured from spectra extracted from 35$-$40 arcsec
radius regions. The count rates measured from spectra were corrected
for the telescope vignetting and PSF. We found that the count rates
obtained using the WT algorithm are ~$\sim$8$\%$ and ~$\sim$4$\%$
lower than the count rates measured from the spectra in the F band and
S band respectively. We then corrected the count rates evaluated using
the WT algorithm by these amounts. The weighted mean of the best
fit spectral indices is $\alpha_E=0.8\pm0.2$ ($1\sigma$ confidence
level). 

\begin{figure}
\begin{center}
\includegraphics[angle=0,height=8truecm,width=8truecm]{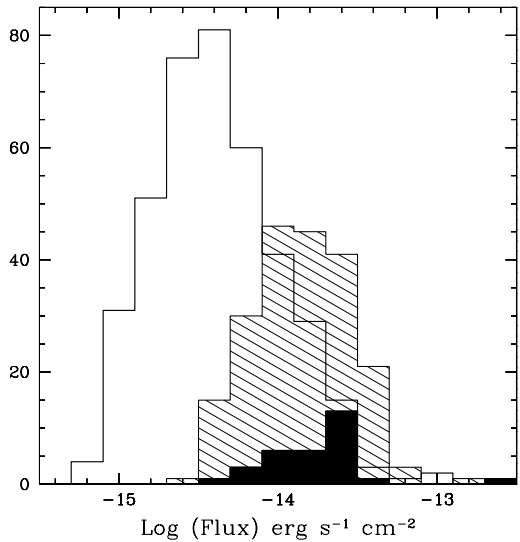}
\caption{The distribution of the detected sources as a function of the
flux in the  0.5-2 keV band (solid
line, empty histogram), 2-10 keV band (shaded histogram) and 5-10 keV
band (black histogram).  }
\label{fxhisto}
\end{center}
\end{figure}

Count rates were converted to fluxes using the following conversion
factors: 1.6$\times10^{-12}$ \cgs (S band), 8.3$\times10^{-12}$ \cgs
(H band) and 16.5$\times10^{-12}$ \cgs (HH band), which are
appropriate for a power law spectrum with energy index
$\alpha_E=0.8$. These three conversion factors are not strongly
sensitive to the spectral shape, due to the narrow bands: for
$\alpha_E=0.4$ and 1.0 they change by up to $\sim1\%$ in the S band
and by up to 10\% in the H band.  For an absorbed power law spectrum
with $N_H=10^{22}$ cm$^{-2}$ and $\alpha_E=1$ the conversion factors
change by $<1\%$ in the H band and by $\sim14\%$ in the S band.  The
conversion factor for the F band depends more on the spectral shape
because of the wider band. For a power law spectrum with $\alpha_E=0.8
\pm_{0.4}^{0.2}$ it changes by up to $30\%$. To minimize this error,
we therefore decided to follow a different approach for the F band: we
used a variable conversion factor which is a function of the spectral
shape, estimated from the hardness ratios (H-S)/(H+S).  The histogram
of the number of sources as a function of the flux is given in
Figure \ref{fxhisto}. We detect sources in the 0.5-2 keV and 2-10 keV
bands down to flux limits of $\sim5.5\times 10^{-16}$ \cgs and
$\sim2\times 10^{-15}$ \cgs respectively.

The ``Sky coverage'' defines the area of the sky covered down to a
given flux limit, as a function of the flux. Due to the telescope
vignetting and to the increase of the size of the PSF in the outer
regions of the detector, the sensitivity decreases toward the outer
detector regions.

\begin{figure}
\begin{center}
\includegraphics[angle=0,height=8truecm,width=8truecm]{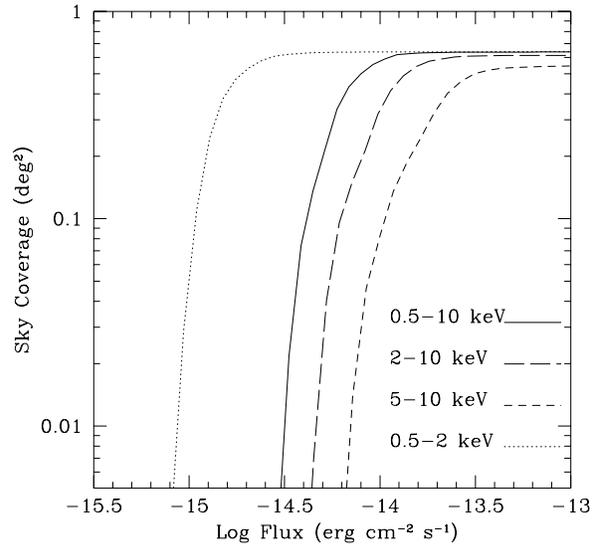}
\caption{The sky coverage in the four 0.5-10 keV, 0.5-2 keV, 2-10 keV
and 5-10 keV bands. The count rate--flux conversion factors are
appropriate for a power law spectrum with $\alpha_E=0.8$. }
\label{skycov}
\end{center}
\end{figure}

The minimum number of counts needed to exceed the Poisson fluctuations
of the background, with a probability threshold of 2$\times$10$^{-5}$
in each detector position has been evaluated using the background maps
in each band and the PSF shape as a function of the off-axis angle.
This minimum number of counts has been divided by the exposure map and
thus converted to minimum detectable fluxes (limiting flux) using the
above defined count rate--flux conversion factors for the S, H and HH
bands. The F band limiting count rates were converted to fluxes using
a conversion factor of 3.25$\times10^{-12}$ \cgs, which is appropriate
for a power law spectrum with $\alpha_E=0.8$. The sky coverage at a
given flux is then obtained by adding up the contribution of all
detector regions with a given flux limit.  Figure \ref{skycov} plots
the resulting sky-coverage in the four bands. In each band the sky
coverage curve is very steep near the faintest detectable flux,
typically decreasing by an order of magnitude for a change in the
limiting flux of $\sim 10 \%$.

The main sky coverage uncertainty is due to the unknown spectrum of
the sources near the detection limit. Of course, this uncertainty
increases with the width of the band and therefore is largest in the
F band.  To estimate, at least roughly, this uncertainty, we
calculated the sky coverage also for power law spectra with
$\alpha_E=0.4$, $\alpha_E=1.0$, and for absorbed power law spectra with
$\alpha_E=0.4,1.0$ and N$_H=10^{22}$ cm$^{-2}$, in addition to the
baseline case.  At fluxes higher than a few $\times 10^{-14}$ \cgs the
percentage deviations of the sky coverage with respect to the baseline
case are negligible for all energy bands.  At a flux of $10^{-14}$
\cgs the deviations are $\sim 75\%$ in the F band, $\sim 30\%$ in the H
band and negligible for the other bands. At a flux of $5\times
10^{-15}$ \cgs the deviations are $\sim 100\%$ in the F band, $\sim 80\%$
in the H band and negligible in the S and HH bands.

\subsection{Source confusion}

Source confusion can affect at some level deep XMM-Newton images.  At
0.5-2 keV and 2-10 keV fluxes of $1.5\times 10^{-15}$ \cgs, and
$5\times 10^{-15}$ \cgs, corresponding to a sky coverage of $\sim$0.04
square degrees, we expect $\sim 750-600$ sources deg$^{-2}$, using the
number counts from the compilation of Moretti et al. (2003). The probability
of finding two sources with the above fluxes within 12'' arcsec from
each other (twice the size of the typical detection cell) is only
$\sim 2-2.5\%$ for the S and H bands, respectively. These
probabilities go up to $\sim(4- 5)\%$ for fluxes corresponding to the
faintest detected sources in the two bands.  The spatial extension and
asymmetry in the count distribution of the sources were checked
interactively, to understand whether same of the X--ray sources can
actually result from the blend of two or more sources. We find that
this may be the case for $\sim14$ sources (i.e., 3\% of the whole F
sample) at fluxes $\ls3\times 10^{-14}$ \cgs and for 2 sources at
fluxes $\sim1.4\times 10^{-13}$ \cgs. Nine of these sources have been
deblended and their count rates has been measured interactively from
the images. The remaining 7 sources have morphologies and surface
brightnesses typical of extended sources (2 of them are bright
clusters of galaxies, see Sect. 2.5). We conclude that source
confusion does not significantly affect our results, especially the
X-ray number counts presented in Sect. 3.

\subsection{Extended sources}

\begin{figure*}[t!]
\begin{center}
\begin{tabular}{cc}
\includegraphics[height=4.5truecm,width=6.5truecm,angle=0]{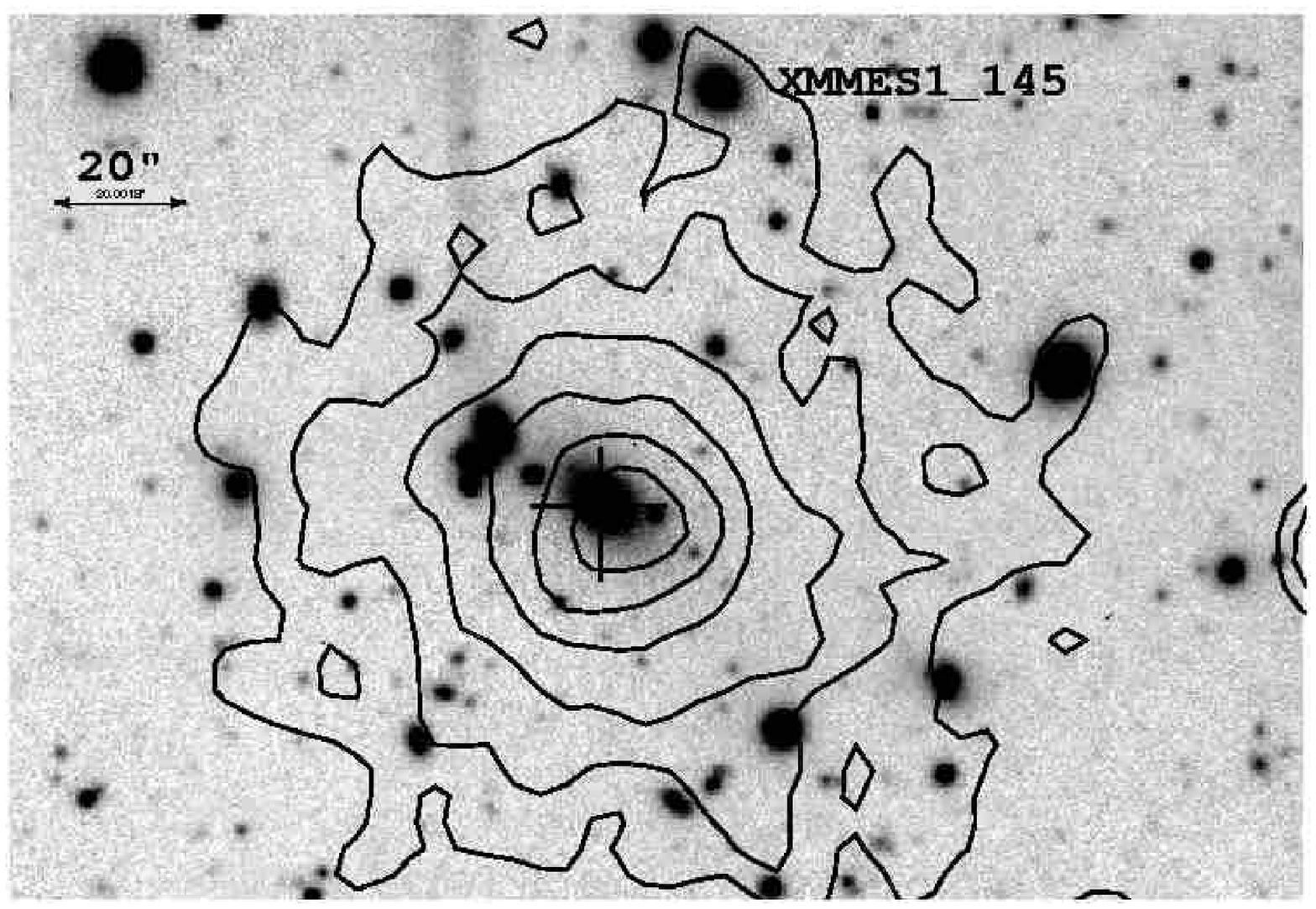}
\includegraphics[height=4.5truecm,width=6.5truecm,angle=0]{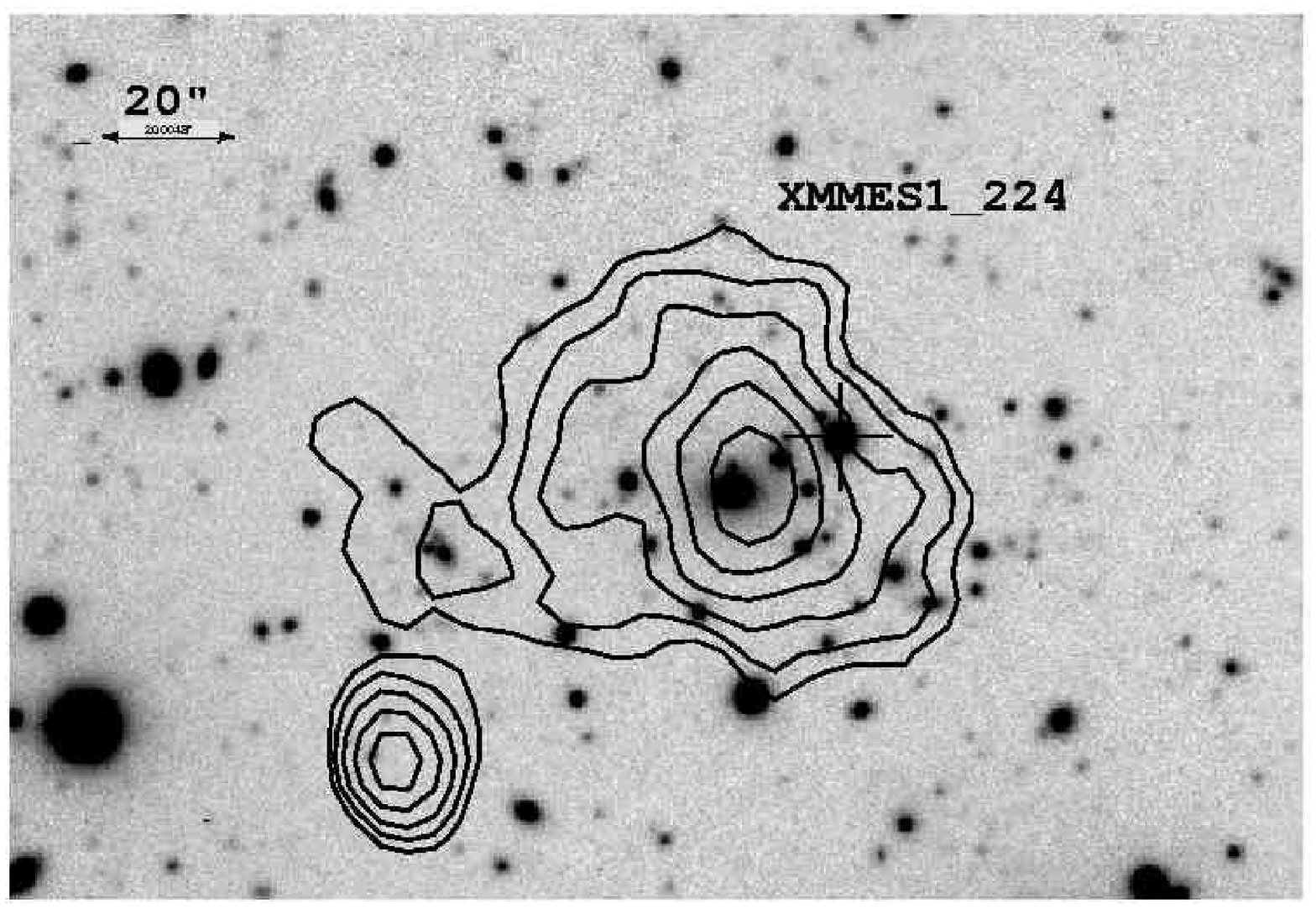}
\end{tabular}
\begin{tabular}{cc}
\includegraphics[height=4.5truecm,width=6.5truecm,angle=0]{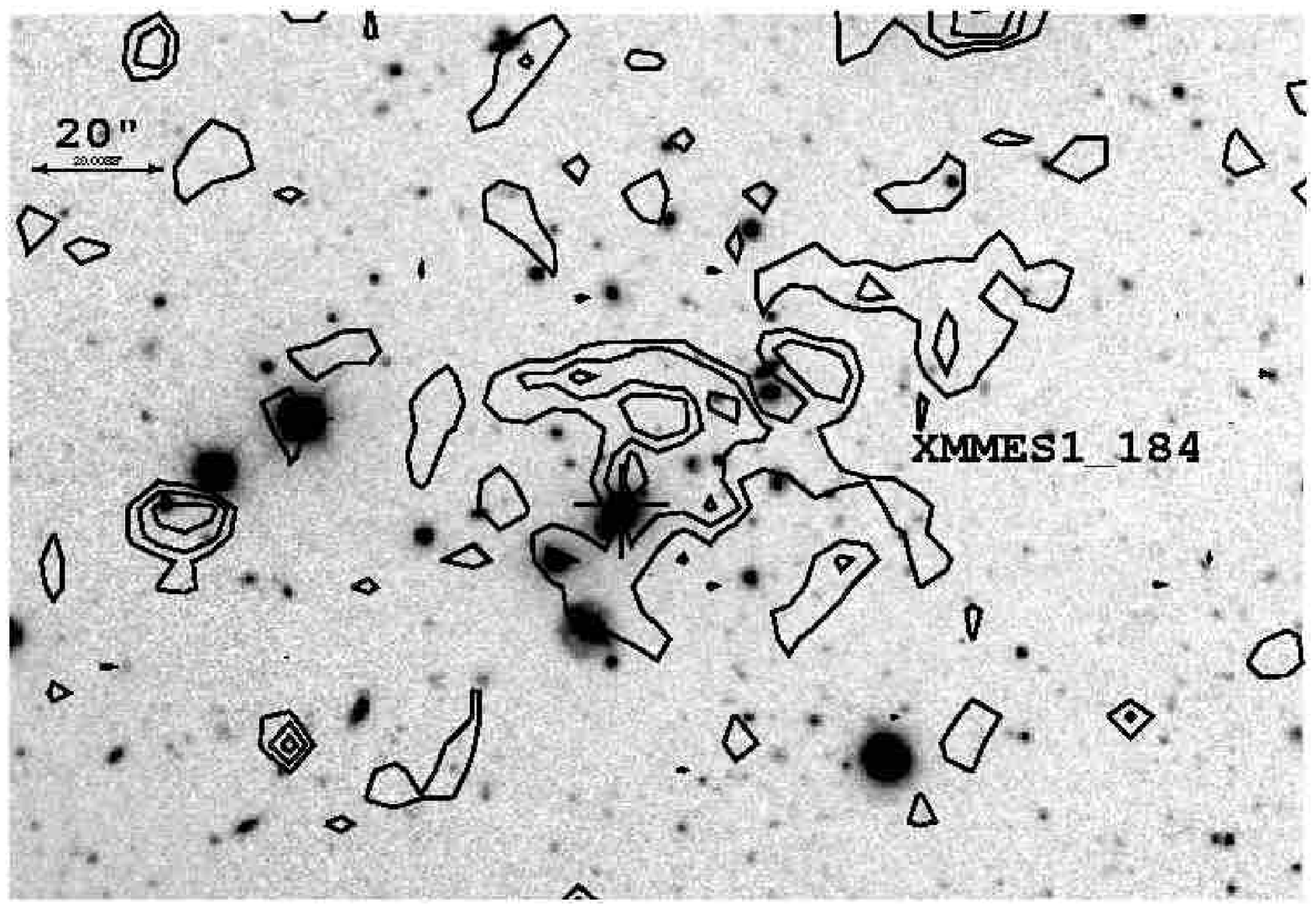}
\includegraphics[height=4.5truecm,width=6.5truecm,angle=0]{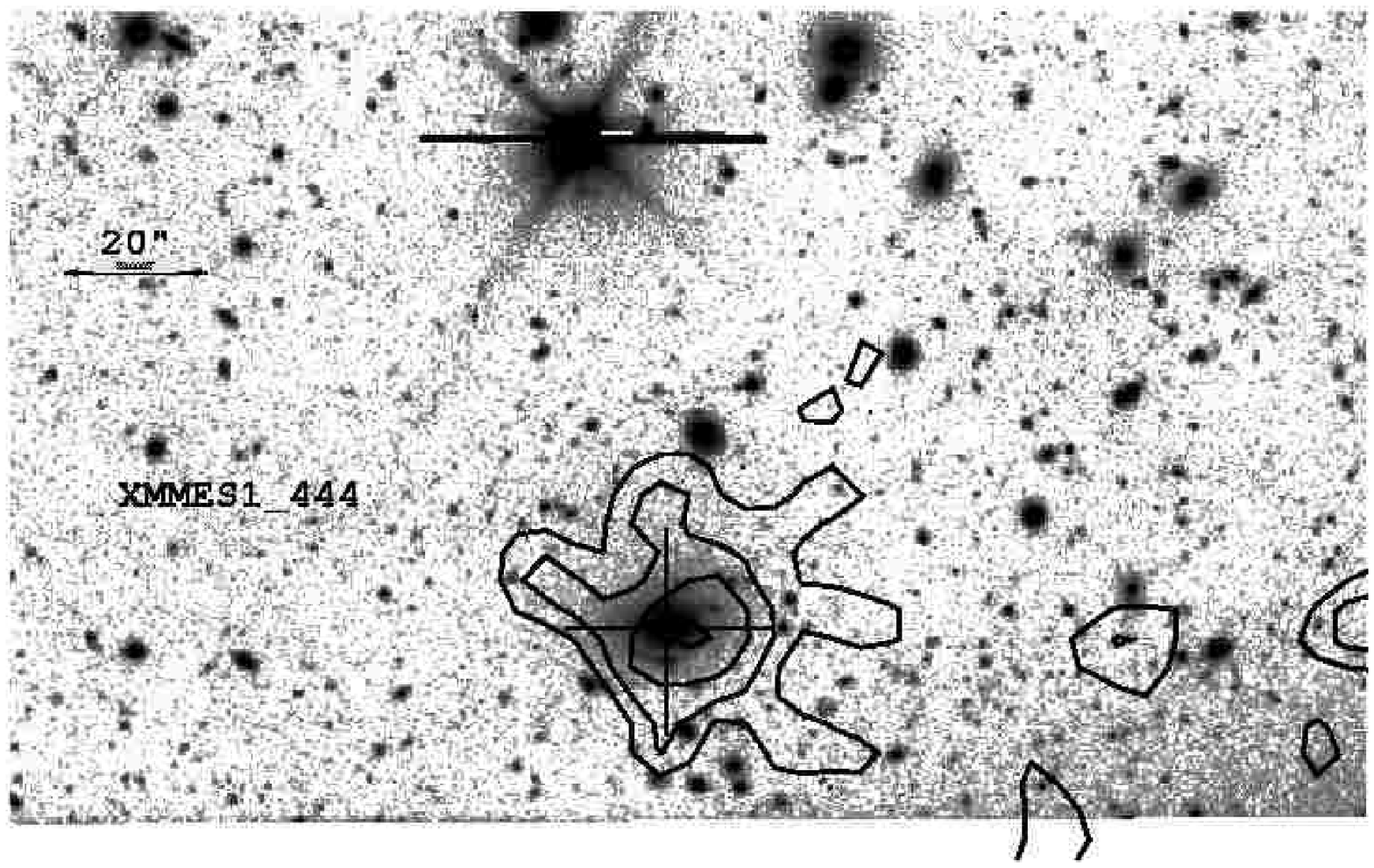}
\end{tabular}
\begin{tabular}{cc}
\includegraphics[height=4.5truecm,width=6.5truecm,angle=0]{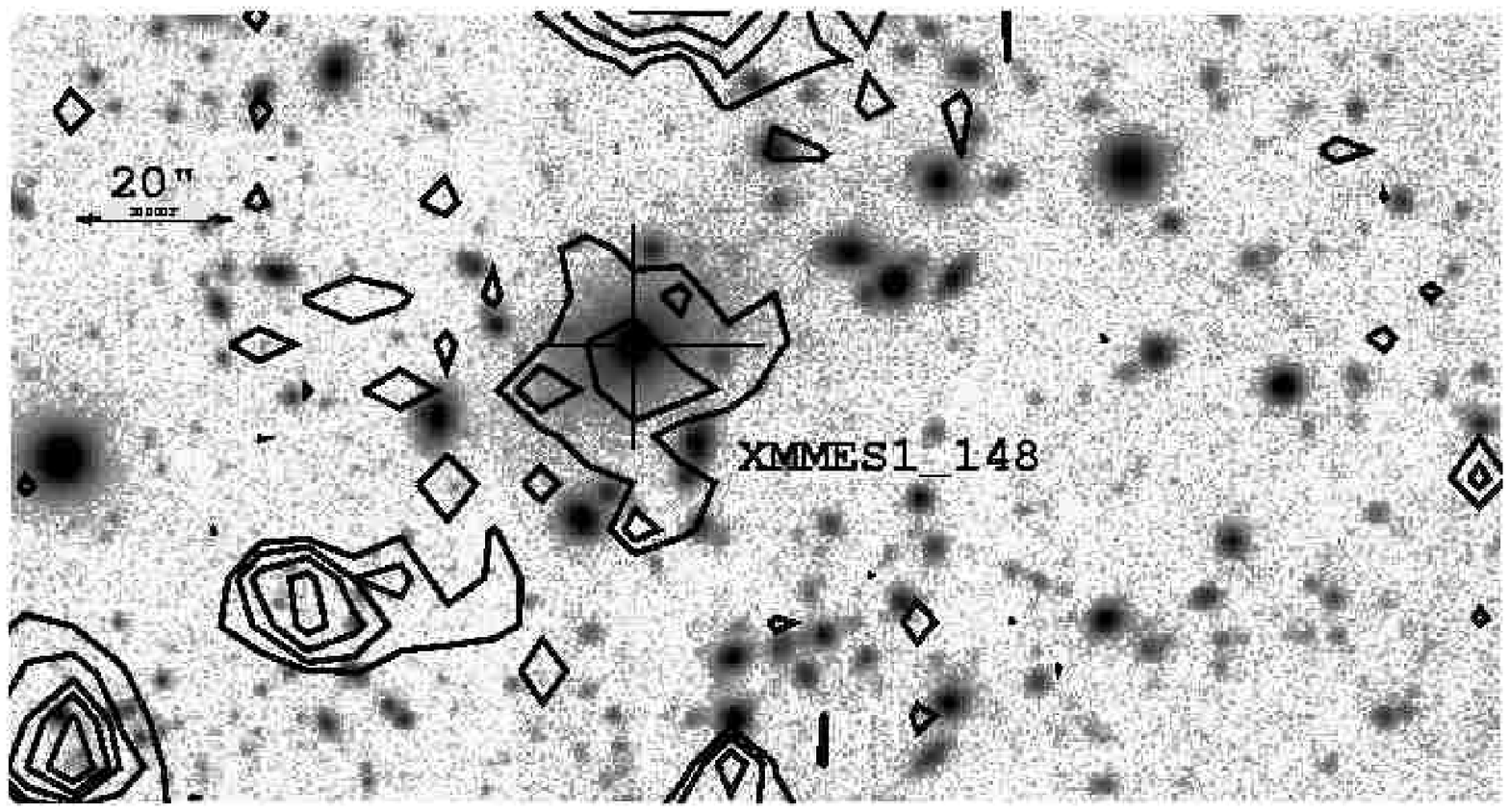}
\includegraphics[height=4.5truecm,width=6.5truecm,angle=0]{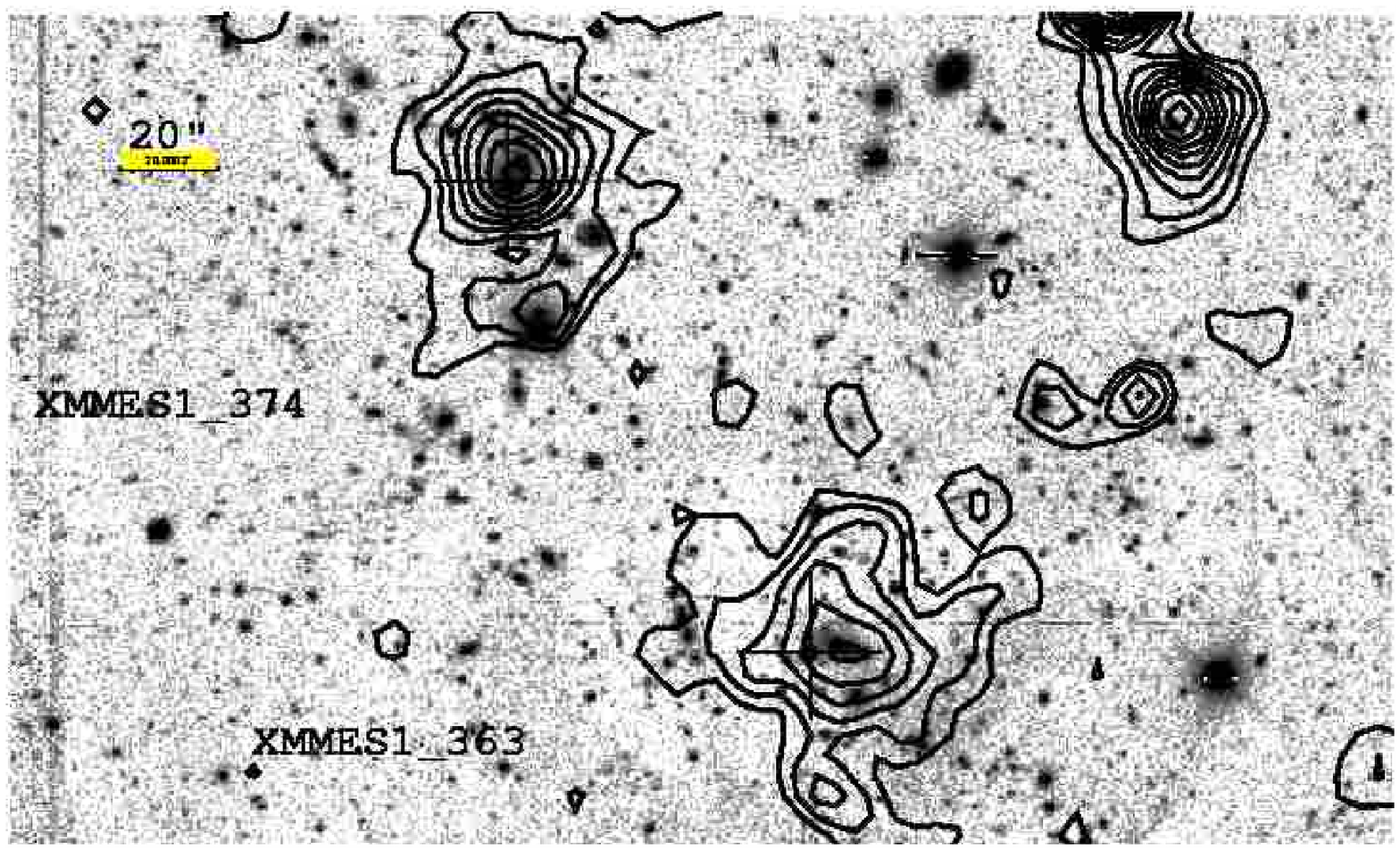}
\end{tabular}
\caption{0.5-10 keV contours of the seven extended sources overlayed on the R
band VIMOS images (XMMES1\_145, XMMES1\_224 and XMMES1\_184, XMMES1\_444,
XMMES1\_374, XMMES1\_363,XMMES1\_148). 
The crosses indicate the brightest galaxy associated with each cluster. }
\label{clusterima}
\end{center}
\end{figure*}

\begin{table*}[ht!]
\begin{center}
\caption{\bf Extended sources, X-ray data}
\begin{tabular}{lcccccccc}
\hline
Source name & R.A. & DEC. & PN$^a$ & MOS$^b$  & F(0.5-10keV) & KT & z & logL$_X$ \\
& J2000 & J2000 & net counts &net counts &$10^{-14}$ c.g.s. & keV & & erg s$^{-1}$ \\
\hline
XMMES1\_145 & 8.44298 & -43.29217 &3228$\pm$74 &  2713$\pm$68 & 14.6 &2.7$\pm$0.2 &
0.21$\pm$0.01 & 43.3 \\
XMMES1\_148 &   8.44469 & -43.34471 &134$\pm$16 &  105$\pm$17  & 0.6  &1.8$\pm_{0.4}^{0.8}$ & 
0.27$\pm_{0.08}^{0.14}$ & 42.2 \\
XMMES1\_184 & 8.52914 & -43.37431 &203$\pm$25 & 155$\pm$22 &  0.8 &1.3$\pm$0.2 &
0.4$\pm_{0.1}^{1.3}$ & 42.8 \\
XMMES1\_224 & 8.61410 & -43.31646 &2448$\pm$58 & 1865$\pm$51 &   14.3 &3.2$\pm$0.3 &
0.39$\pm$0.02 & 43.9 \\
XMMES1\_363 & 8.94811 & -43.38005 &303$\pm$25 &  261$\pm$23  &1.6 &2.7$\pm_{0.5}^{0.7}$ & 
1.10$\pm$0.13& 44.1 \\
XMMES1\_374 & 8.97218 & -43.35345 &283$\pm$24 &   224$\pm$22   &  1.2 &1.1$\pm$0.1 &
0.37$\pm$0.06 & 42.7 \\
XMMES1\_444 & 9.11833 & -43.47706 &201$\pm$22 & 176$\pm$20 & 2.5 & 3.1$\pm_{1.1}^{2.5}$ &
0.7$\pm_{0.4}^{3.3}$ & 43.7 \\
\hline   

\end{tabular}
\end{center}

$^a$Net PN counts in the 0.5-10 keV energy range.
$^b$Net MOS1+MOS2 counts in the 0.3-10 keV energy range.
The quoted errors corresponds to 68\% confidence level.
\end{table*}

\begin{table*}[ht!]
\begin{center}
\caption{\bf Extended sources: optical and near infrared data$^1$ of the 
  brightest galaxy associated with each cluster.}
\begin{tabular}{lcccccccc}
\hline
Source name & R.A. & DEC. & B     & V     & R     & K     &J& z$_{phot}$ \\ 
            & J2000 & J2000 & mag & mag & mag & mag & mag & \\
\hline
XMMES1\_145 & 8.4433322 & -43.291992  &  19.24 & 17.76 & 16.99 &  13.86 &15.15 &   0.15-0.30 \\
XMMES1\_148 & 8.4452075 & -43.344633  &  20.07 & 18.54 & 17.74 &  14.57 &15.90 &  0.15-0.75 \\
XMMES1\_184 & 8.5332721 & -43.377095  &  20.79 & 19.31 & 18.25 &  14.59 &16.13 &  0.25-0.80 \\
XMMES1\_224 & 8.6080313 & -43.314348  &  20.79 & 19.19 & 17.93 &  14.36 &14.55  &   0.25-0.85 \\
XMMES1\_363 & 8.948075  & -43.379401  &  23.31 & 21.34 & 20.21 &  16.13 &17.48 &   0.65-0.95 \\
XMMES1\_374 & 8.9731188 & -43.353933  &  22.23 & 20.56 & 19.47 &  16.07 &17.41 &   0.25-0.80 \\
XMMES1\_444 & 9.1195568 & -43.475508  &  21.07 & 19.44 & 18.24 &  14.63 &16.30 &   0.30-0.80 \\
\hline   

\end{tabular}
\end{center}

$^1$ Magnitude in the Johson-Cousins Vega system.

\end{table*}

\begin{figure*}
\begin{center}
\begin{tabular}{cc}
\includegraphics[angle=0,height=6.2truecm,width=6truecm]{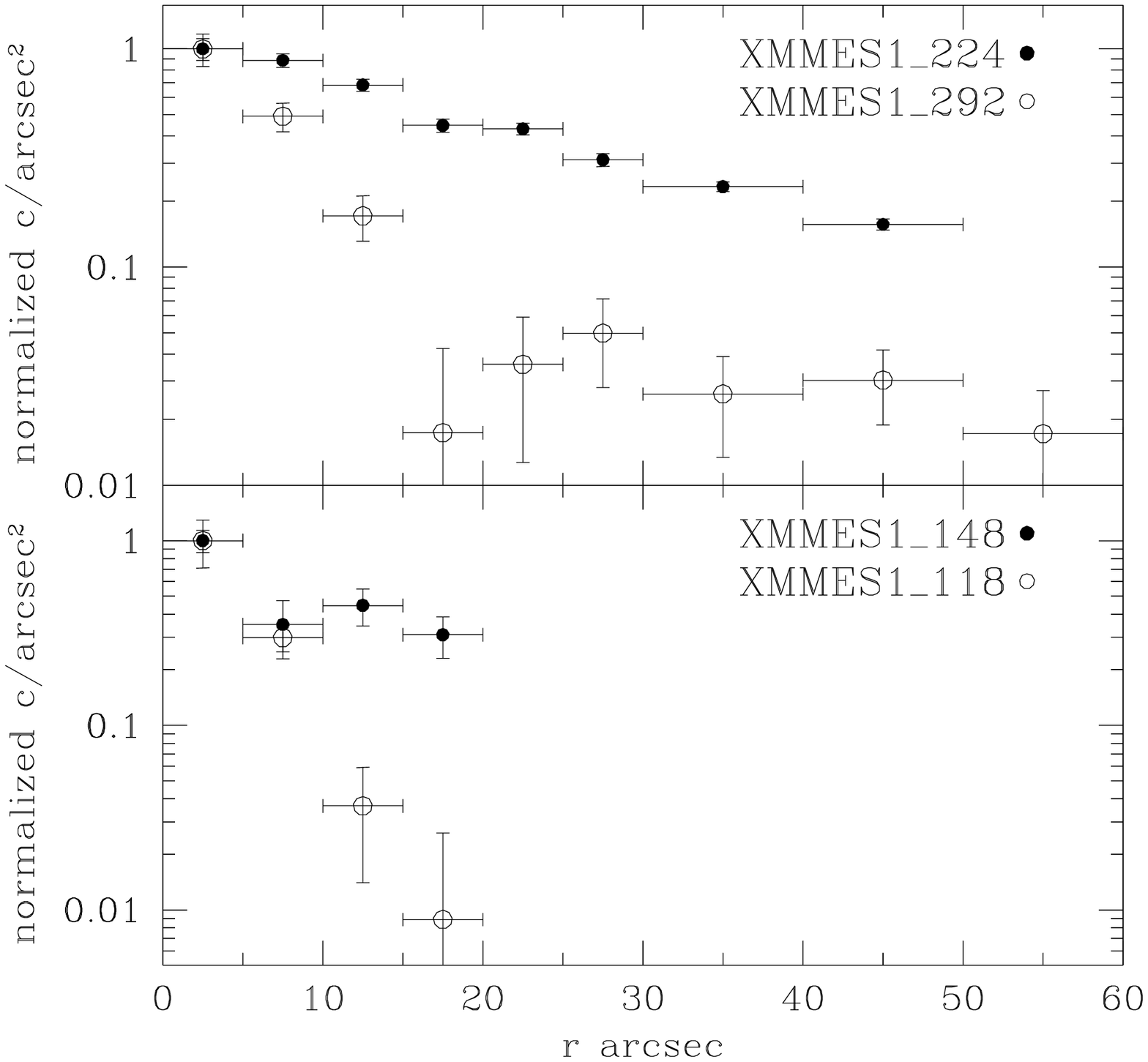}
\includegraphics[angle=0,height=6.2truecm,width=6truecm]{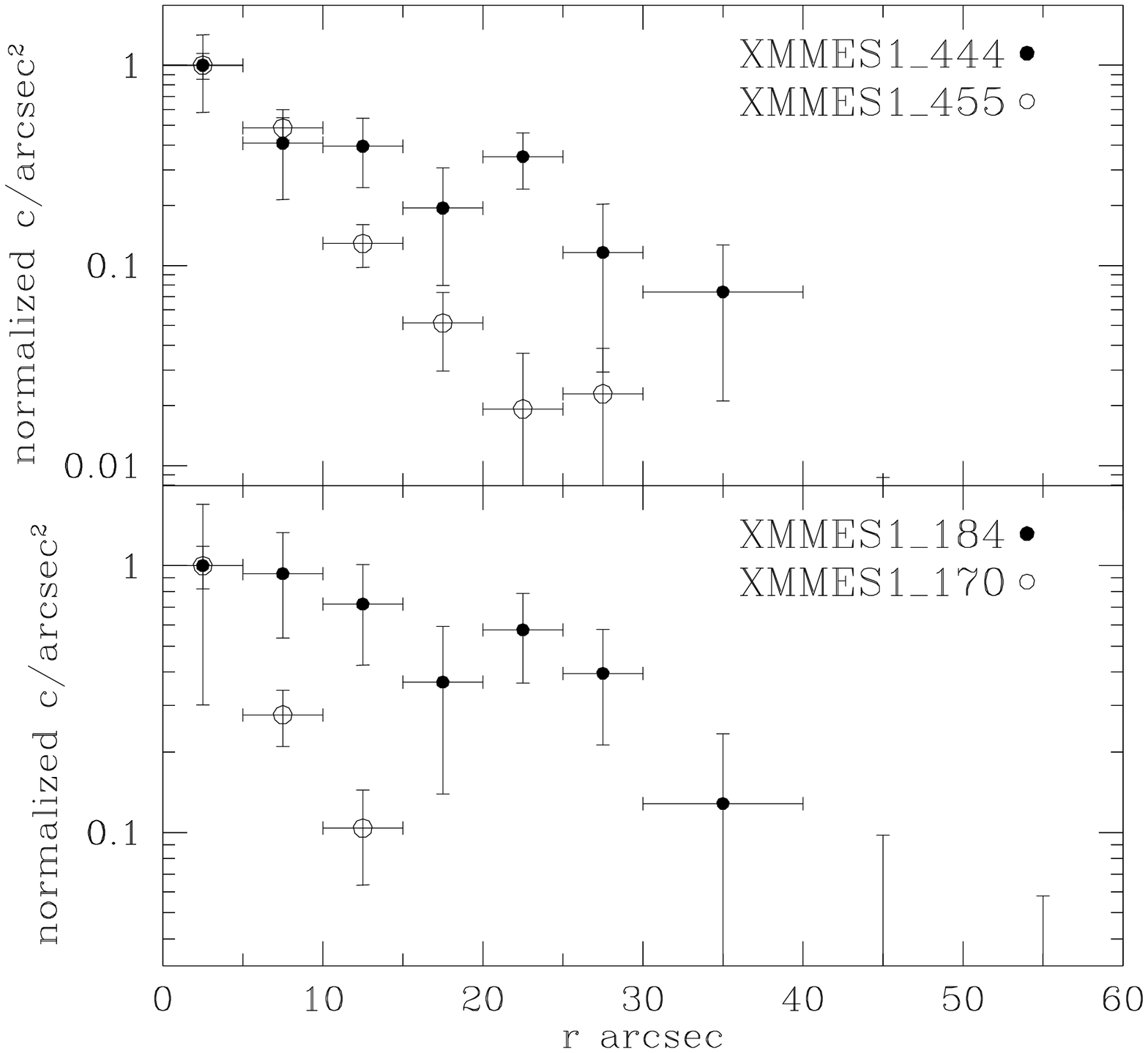}
\end{tabular}
\begin{tabular}{cc}
\includegraphics[angle=0,height=6.2truecm,width=6truecm]{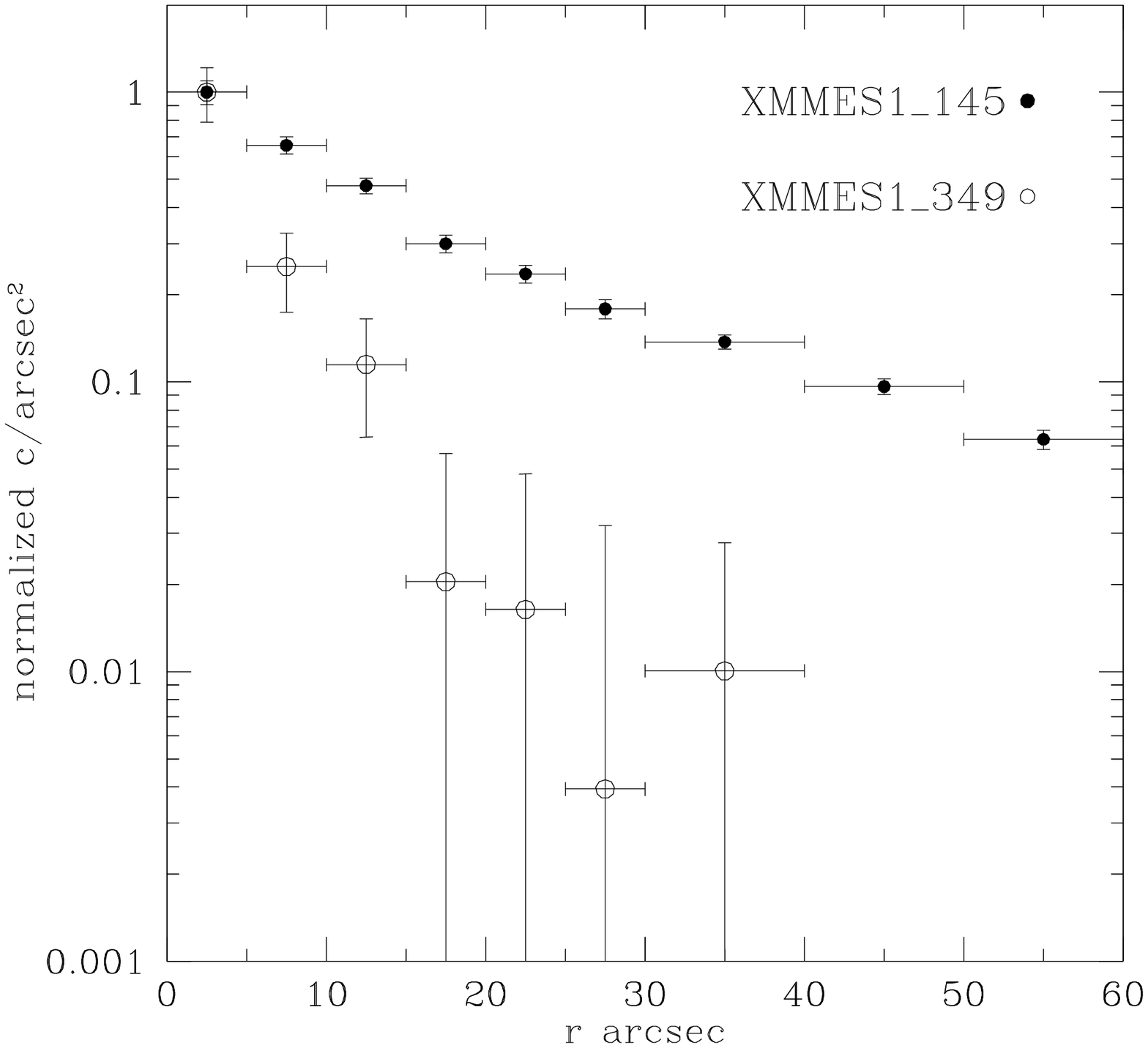}
\includegraphics[angle=0,height=6.2truecm,width=6truecm]{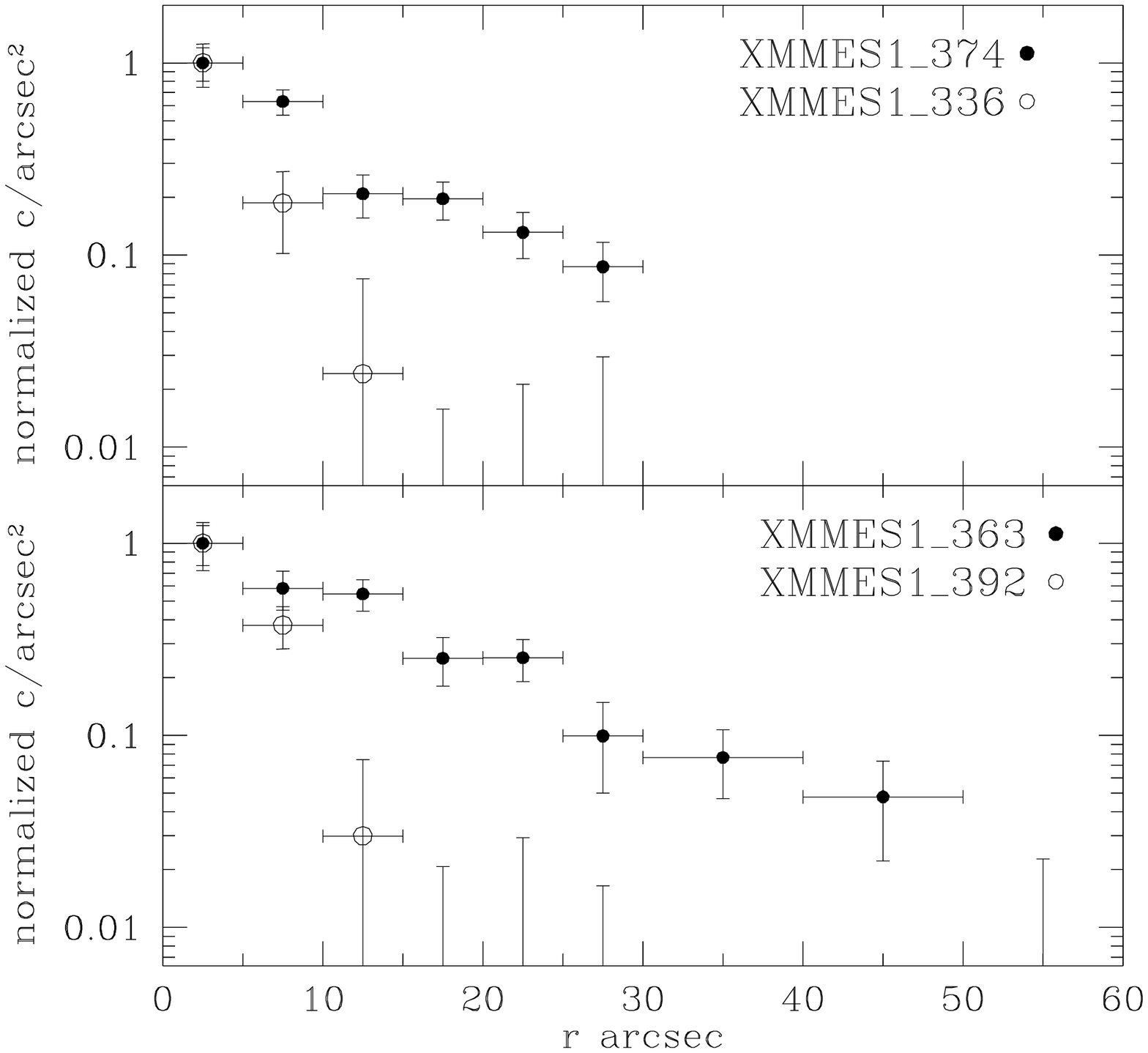}
\end{tabular}
\caption{Radial distribution of the background subtracted surface
brightness, normalized to the peak central value, of the sources
XMMES1\_145, XMMES1\_148, XMMES1\_184, XMMES1\_224, XMMES1\_363,
XMMES1\_374 and XMMES1\_444, computed in concentric annuli (full
dots), compared that of point--like sources at the same off-axis
angles (open dots).}
\label{surfacebri}
\end{center}
\end{figure*}

We identified seven extended X-ray sources by comparing their
background subtracted surface brightness, computed in concentric
annuli, to the background subtracted surface brightness of point-like
sources found at similar off-axis angles, which provide a reasonably
good representation of the local PSF. The surface brightness radial
profiles of these sources, normalized to the peak surface brightness
are shown in Figure \ref{surfacebri}. For all of them, the difference
with the radial profiles of the point-like sources are significant at
more than 99.9\%.

Figure \ref{clusterima} shows the X-ray contours of the 7 sources
overlayed on deep R band optical images obtained using VIMOS@VLT ESO
telescope.  We see that all sources can be identified with clusters or
groups of galaxies.

\begin{figure*}
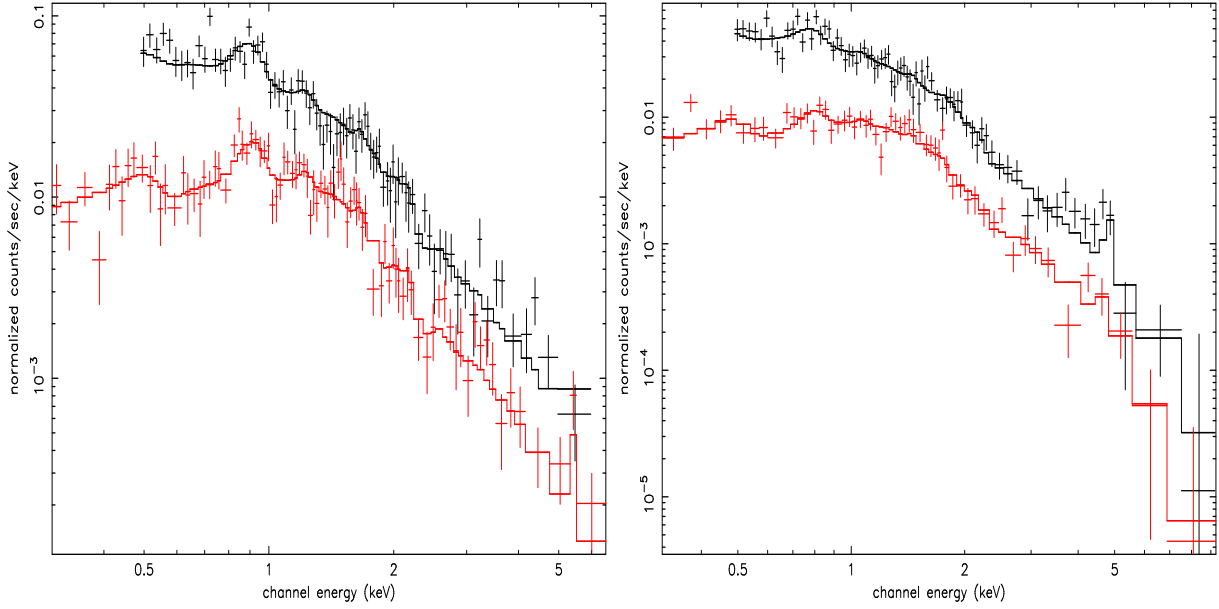

\begin{center}
\begin{tabular}{cc}
\includegraphics[height=8truecm,width=8truecm,angle=270]{cluster_1_reb315_setreb2800.ps}
\includegraphics[height=8truecm,width=8truecm,angle=270]{cluster_2_reb320_setreb1800.ps}
\end{tabular}
\caption{PN and MOS counts spectra of the two brighest extended
sources, XMMES1\_145 (left panel) and XMMES1\_224 (right panel), along
with their best fit thermal models.}
\label{spettri}
\end{center}
\end{figure*}

\begin{figure*}
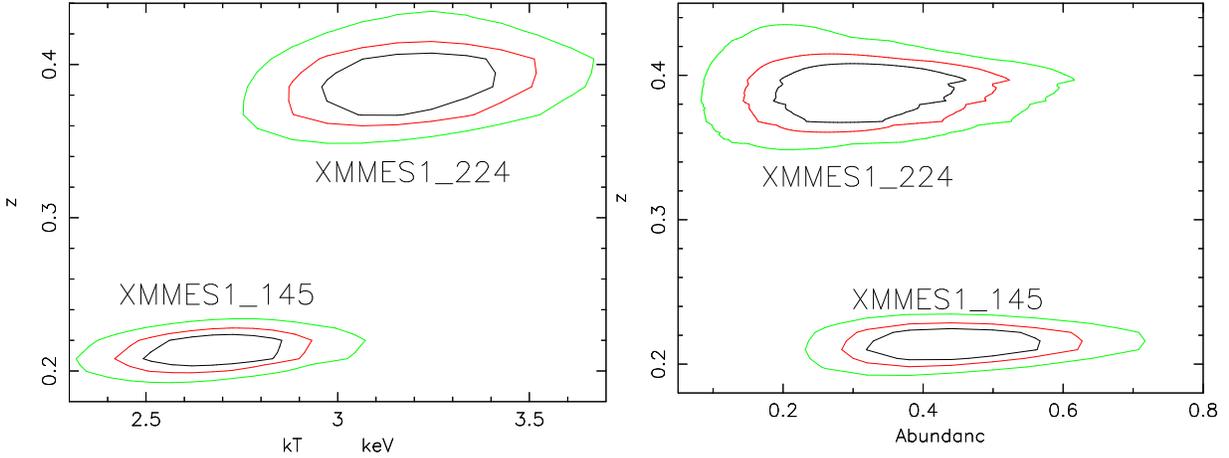

\begin{center}
\begin{tabular}{cc}
\includegraphics[angle=-90,width=8truecm]{2clu_chi_zt.ps}
\includegraphics[angle=-90,width=8truecm]{2clu_chi_abz.ps}
\end{tabular}
\caption{Left panel: the 68\%, 90\% and 99\% redshift-temperature
$\chi^2$ contours for the two clusters of galaxies XMMES1\_145 and
XMMES1\_224. Right panel: The 68\%, 90\% and 99\% redshift-metal
abundance $\chi^2$ contours.}
\label{ztclu}
\end{center}
\end{figure*}

We extracted the X-ray spectra of the 7 extended sources from circular
regions with radii in the range 35$\arcsec$ to 70$\arcsec$. The
background counts were extracted from the nearby source-free regions.
The response and ancillary files were generated by the XMM-SAS tasks,
{\sc rmfgen} and {\sc arfgen}, respectively. The MOS1 and MOS2 spectra
are combined together. The spectra were fitted using XSPEC (version
11.3.1, \cite{arnaud03}), to a thin plasma thermal model with free
temperature and redshift.  For the two brightest sources (XMMES1\_145
and XMMES1\_224, see Table 4), the statistics is good enough to let
also the metal abundance to be free to vary in the fit. Figure
\ref{spettri} shows the PN and MOS spectra of these two sources along
with the best fit thermal model. In the other five cases, we fixed the
metal abundance to the typical value (0.3) found in clusters of
galaxies over broad redshift range (see, e.g., \cite{tozzi}). Table 4
gives for each of the extended sources the X-ray position, the net
counts in the 0.5-10 keV energy range and the corresponding flux, the
best fit temperature, redshift and X--ray luminosity.  

\begin{figure}
\begin{center}
\begin{tabular}{cc}
\includegraphics[angle=-90,width=8truecm]{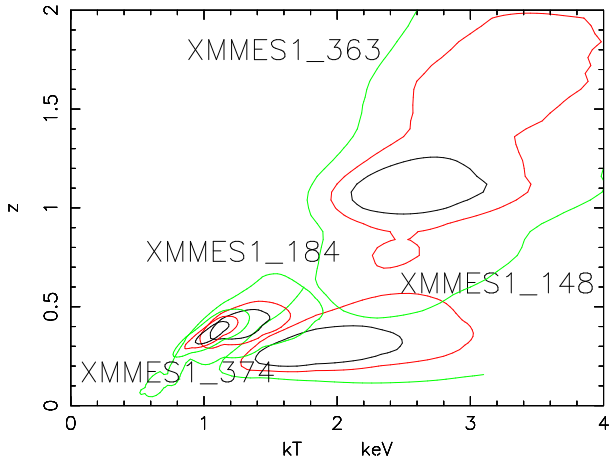}
\end{tabular}
\caption{The 68\%, 90\% and 99\% redshift-temperature $\chi^2$
contours for the sources XMMES1\_148, XMMES1\_184, XMMES1\_363 and
XMMES1\_374.}
\label{zt4}
\end{center}
\end{figure}

Figure \ref{ztclu} shows the temperature-redshift and redshift-metal
abundance $\chi^2$ confidence contours for the sources XMMES1\_145 and
XMMES1\_224.  The quality of the XMM-Newton data is such to tightly
constrain the redshift of the two bright clusters (0.21$\pm$0.01 and
0.39$\pm$0.02). Figure \ref{zt4} shows the temperature-redshift
$\chi^2$ contours for the sources XMMES1\_374, XMMES1\_363,
XMMES1\_184 and XMMES1\_148.  Although in these cases the statistics
is limited (see counts and fluxes in Table 4) the XMM-Newton data are
still good enough to provide rough information on the redshift of the
sources. Note that the redshift of XMMES1\_363 is constrained to be
$>0.7$ at the 90\% confidence level, and that 4 of the sources (
XMMES1\_224, XMMES1\_374, XMMES1\_184 and XMMES1\_148) are consistent
with having the same redshift (z$\sim$0.4).  The luminosities and
temperatures in Table 4, both determined through the X-ray spectral
fittings, are in good agreement with the luminosity-temperature
relationship of groups and clusters of galaxies (see, e.g.,
\cite{arnaud} and references therein).

Table 5 gives the position and
the B, V, R, K and J magnitudes of the brightest galaxies of each of
the seven X-ray sources in Figure \ref{clusterima}.  We estimated their
photometric redshifts by comparing the observed optical to near
infrared spectral energy distribution with the Rocca-Volmerage galaxy
templates (see, e.g., Fioc \& Rocca-Volmerage 1997). Unfortunately, the
available photometry does not allow a precise determination of the
redshifts of the galaxies, but the redshift confidence intervals in
Table 5 are in all cases consistent with the redshift estimates in
Table 4. In particular, the photometric redshifts confirm that
XMMES1\_363 is at a redshift higher than the rest of the extended
sources sample.

\section{The X-ray number counts}

\begin{figure*}[ht!]
\begin{center}
\includegraphics[angle=0,height=12truecm,width=12truecm]{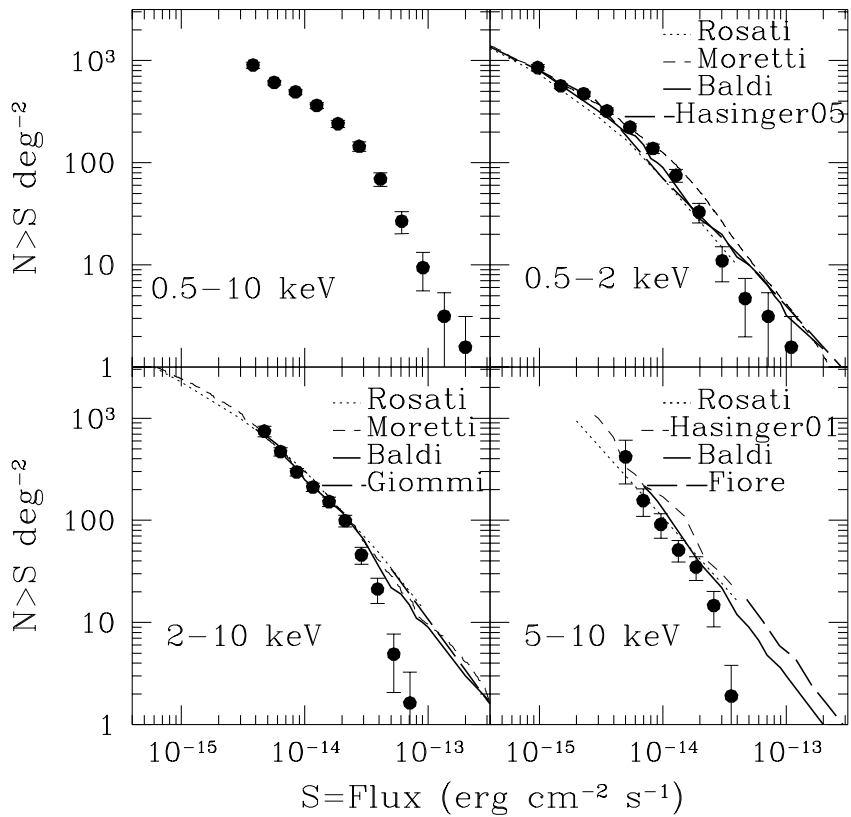}
\caption{The source number counts in the 0.5-10 keV F band (upper left
panel), 0.5-2 keV S band (upper right panel), 2-10 keV H band (lower
left panel), 5-10 keV HH band (lower right panel). The Figure also
shows a few number counts from literature: the 0.5-2 keV number counts
from Rosati et al. (2002), Moretti et al. (2003), Baldi et al. (2002)
and Hasinger et al. 2005 (upper right panel); the 2-10 keV number from
Rosati et al. (2002), Moretti et al. (2003), Baldi et al. (2002) and
Giommi et al. (2000) (lower left panel); the 5-10 keV number from
Rosati et al. (2002), Hasinger et al. (2001), Baldi et al. (2002) and
Fiore et al. (2001) (lower right panel).}
\label{lnls}
\end{center}
\end{figure*}

The XMM-Newton survey in the ELAIS-S1 area covers a range of fluxes
over an area which is large enough to constrain source counts in a
flux range bridging deep, pencil beam, and shallower, large area
surveys.  Figure \ref{lnls} plots the integral number counts in the F,
S, H and HH bands.  Table 6 gives the number counts in the S, H and HH
bands, which can be more easily compared with previous determinations,
together with the sky coverage in the same bands.  Previous
determinations of the S, H and HH number counts are over-plotted on
the ELAIS-S1 number counts. The 0.5-2 keV number counts are compared
to those of Rosati et al. (2002) (Chandra Deep Field South, CDFS),
Moretti et al. (2003) (compilation), Baldi et al. (2002) (HELLAS2XMM)
and Hasinger et al. (2005) (compilation). The CDFS number counts are
the lowest, as already noted by Rosati et al. (2002). The ELAIS-S1
determination is consistent within the errors with those of Moretti et
al. (2003) and Baldi et al. (2002).  The ELAIS-S1 2-10 keV number
counts are compared to those of Rosati et al. (2002), Moretti et
al. (2003), Baldi et al. (2002) and Giommi et al. (2000)
(BeppoSAX). All determinations are consistent with each others at
fluxes lower than $2-3\times10^{-14}$ \cgs.  The ELAIS-S1 5-10 keV
number counts are compared to the those of Rosati et al. (2002),
Hasinger et al. (2001) (Lockman Hole), Baldi et al. (2002) and Fiore
et al. (2001) (BeppoSAX). The ELAIS-S1 number counts are slightly
lower than the other determinations but still consistent within the
errors. In summary, the agreement of the ELAIS-S1 number counts with
other determinations is reasonably good, implying that the ELAIS-S1
region, at least from this point of view, can be considered as
representative of the average X-ray source population.

\begin{table}[ht!]
\caption{\bf Integral number counts and sky coverage}
\begin{tabular}{lccc}
\hline
Flux   & N$^1$& Counts  & Sky-coverage\\
$10^{-14}$\cgs &  & deg$^{-2}$  & deg$^{2}$ \\
\hline
\multicolumn{4}{c}{0.5-2 keV}\\
\hline
0.10   & 23   &       854$\pm$60 & 0.03 \\
0.15    & 50  &     564$\pm$31 & 0.38 \\
0.23    & 61 &     473$\pm$27 & 0.57 \\
0.36    & 84  &   321$\pm$22 & 0.62 \\
0.55    & 59   &     223$\pm$19 & 0.63 \\
0.83    & 48    &     138$\pm$15 & 0.64 \\
1.30    & 30   &     75$\pm$11   &  0.64  \\ 
1.99    & 23   &    33$\pm$7 &  0.64  \\
3.04    & 5    &    11$\pm$4  &  0.64  \\
4.55    & 4    &   5$\pm$3 &  0.64  \\
7.15   & 1&    3$\pm$2& 0.64 \\
11.00 & 1&2$\pm$2&0.64 \\
\hline
\multicolumn{4}{c}{2-10 keV}\\
\hline
0.47   &  18   &    746$\pm$91 &  0.04 \\
0.63    & 19   &    470$\pm$43 &   0.10 \\
0.86    & 26   &      297$\pm$25 &   0.22 \\
1.16    & 32   &      212$\pm$19 &  0.42 \\
1.57    & 35   &      152$\pm$16 &  0.54 \\
2.12    & 27   &     99$\pm$13 & 0.59 \\
2.87   & 22    &    46$\pm$9 & 0.61 \\
3.88    & 15    &     21$\pm$6 & 0.61 \\
5.25    &  4  &     5$\pm$3 & 0.61 \\
7.10 &1& 2$\pm$2 &0.61\\
\hline
\multicolumn{4}{c}{5-10 keV}\\
\hline
0.69   & 3  &    156$\pm$47  &   0.013\\
0.96   & 4  &   91$\pm$24  &   0.07\\
1.34   & 3  &     51$\pm$12  & 0.17\\
1.86   & 7  &     35$\pm$9  &  0.31\\
2.59   & 11  &    15$\pm$5  &  0.45\\
3.60 & 1& 2$\pm$2 & 0.52\\
\hline
\end{tabular}

$^1$ Detected sources per flux bin. Note that a few sources fall out of
the flux range, which was used to evalued the integral number counts in
each energy band.

\end{table}

\section{X-ray spectral properties}

\begin{figure*}[t!]
\begin{center}
\begin{tabular}{cc}
\includegraphics[angle=0,height=8truecm,width=8.5truecm]{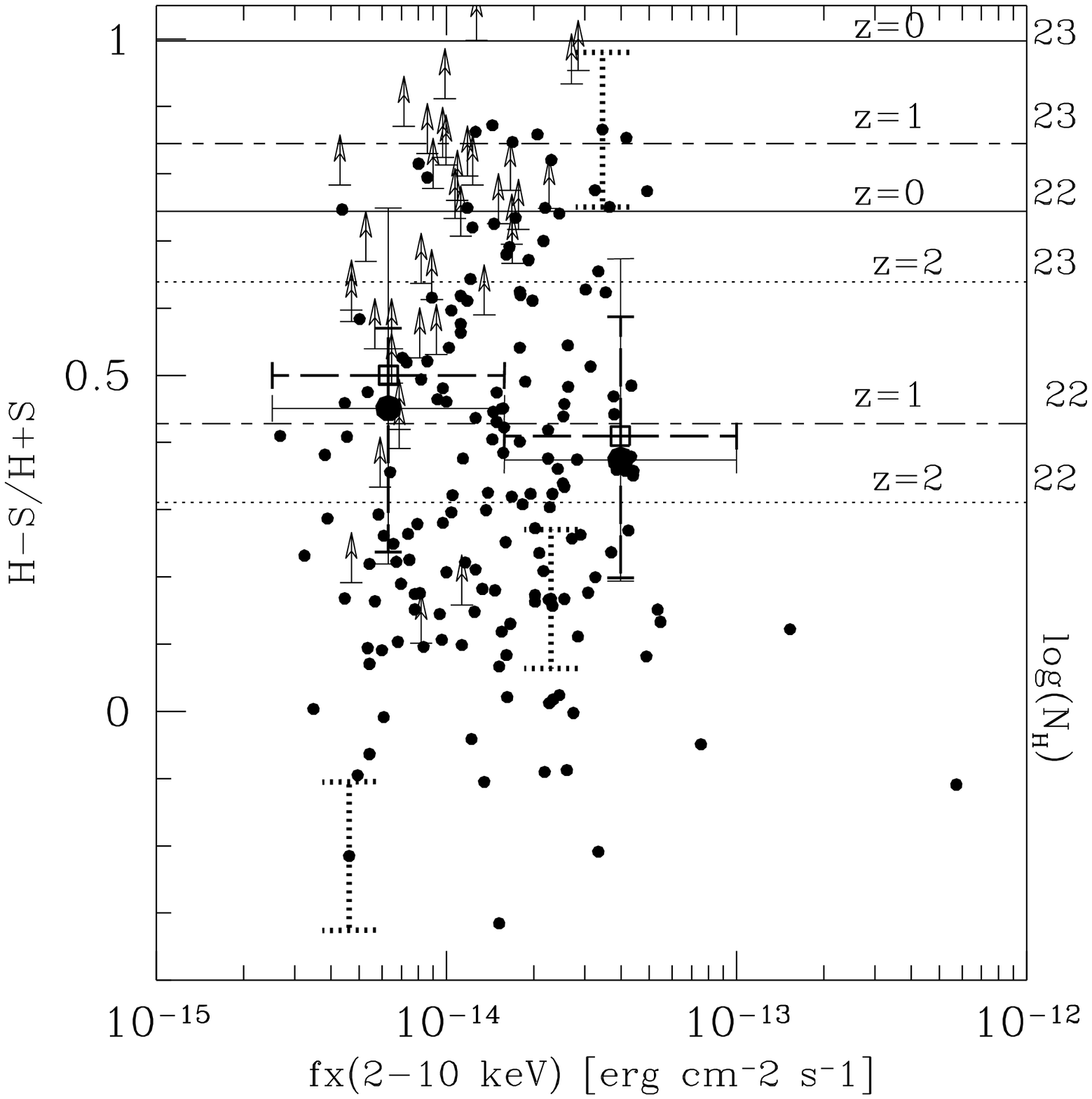}
\includegraphics[angle=0,height=8.2truecm,width=8truecm]{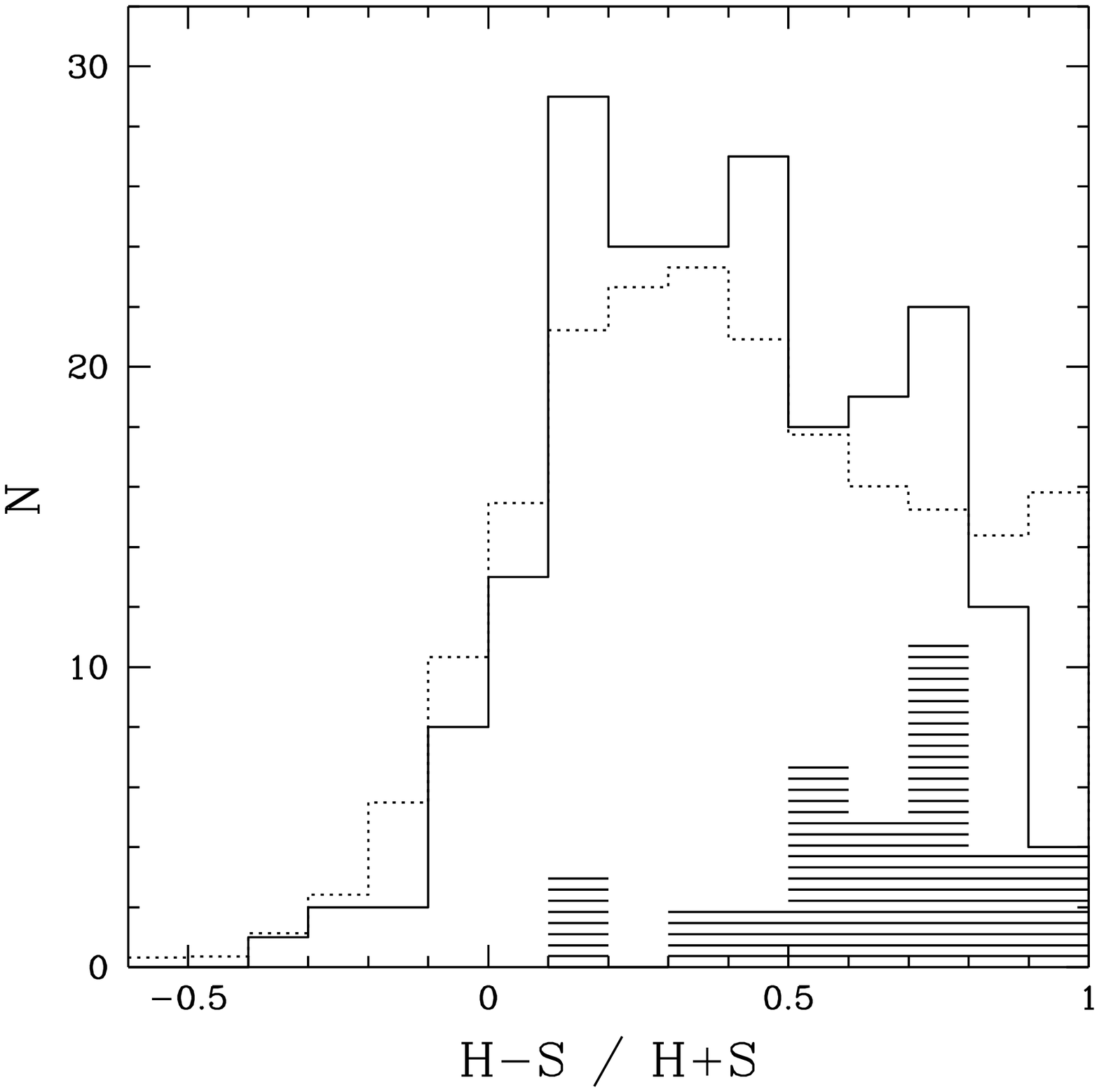}
\end{tabular}
\caption{Left panel: The hardness ratio (H-S)/(H+S) as a function of
the flux for the whole 2-10 keV sample. Typical hardness ratio errors
are shown at different fluxes. 1$\sigma$ upper limits on the S band
flux are shown with by arrows.  Large filled circles represent the
median and interquartile in two flux ranges, evalued using the so
called ``survival analysis''(\cite{kaplan}, \cite{miller}). They are
compared with the same determination from the HELLAS2XMM sample (open
squares). Constant lines are the hardness ratios expected from a power
law model with slope $\alpha=$0.8, reduced at low energy by
photoelectric absorption from neutral gas (N$_H$$=$ 10$^{22}$
cm$^{-2}$ and 10$^{23}$ cm$^{-2}$) at three representative redshifts:
z=0, z=1 and z=2. Right panel: The distribution of (H-S)/(H+S)
for the same sample in the left panel (solid histogram). Upper limits
are identified by the shaded histogram.  The dotted histogram is the
average of the simulated (H-S)/(H+S) distributions.}
\label{hrt}
\end{center}
\end{figure*}

\begin{figure}
\begin{center}
\includegraphics[angle=0,height=8truecm,width=8truecm]{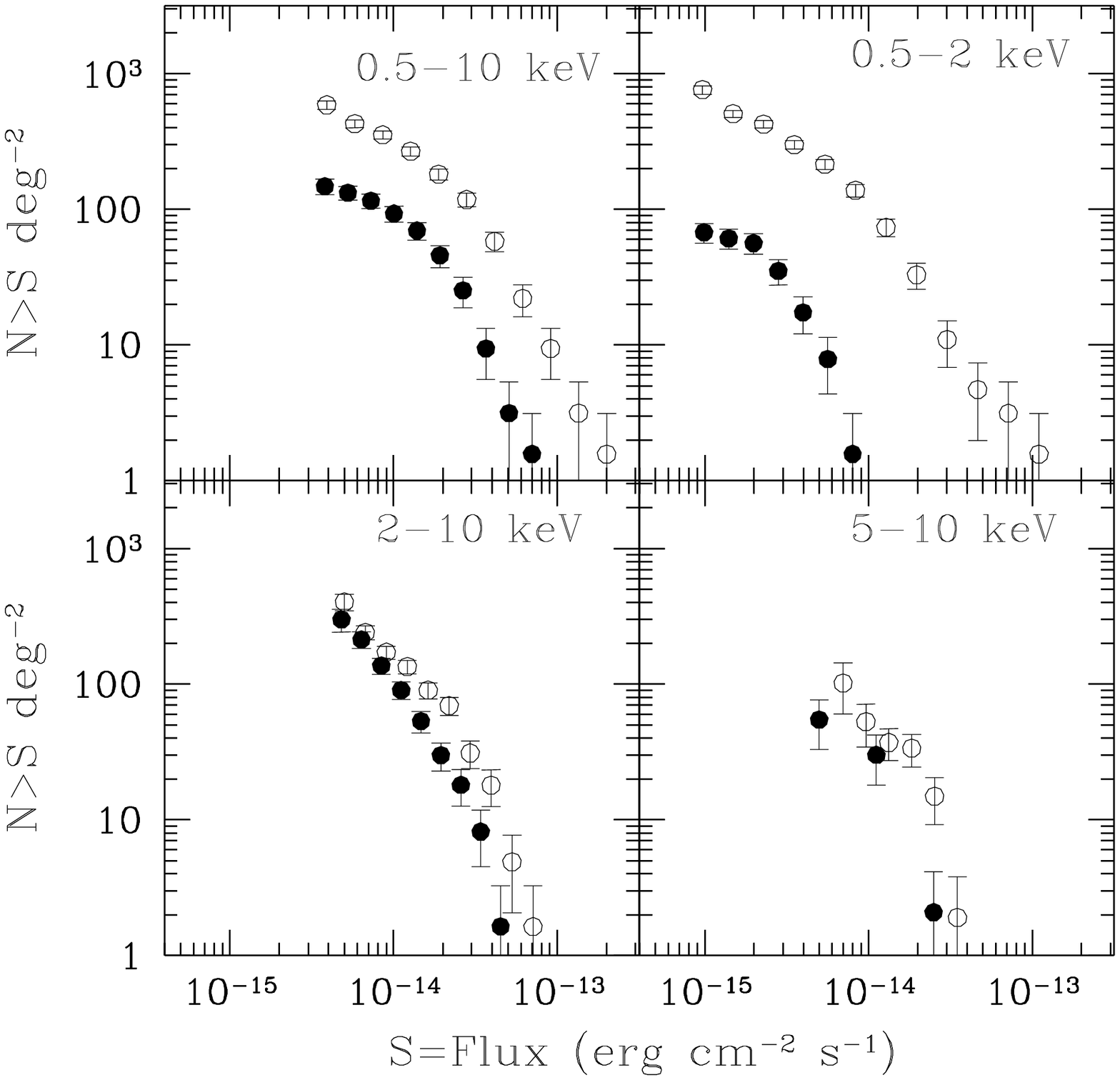}
\caption{The source number counts in the F band (upper left panel), S
band (upper right panel), H band (lower left panel), HH band (lower
right panel). The open circles indicate the source number counts of
the unobscured sources. The solid circles indicate the source number
counts of the obscured sources, as defined in Sect. 4.}
\label{lnlsabuna}
\end{center}
\end{figure}

The broad band X-ray spectral properties of the ELAIS-S1 sources can
be investigated using hardness ratios.  Many of the ELAIS-S1 sources
have a high hardness ratio HR$=$(H-S)/(H+S) (where H and S are the
fluxes in the hard and soft band, respectively, (see, e.g., Fiore et
al. 2003)), indicating a hard spectrum.  H-S/H+S for the full 2-10 keV
sample, 205 sources, is plotted in Figure \ref{hrt}, left panel, as a
function of the 2-10 keV flux.  Lines of constant $N_H$ are overlaid.
Following Hasinger et al. (2003) a threshold of the (H-S)/(H+S) ratio
can be used to roughly separate obscured from unobscured sources. To
this purpose we use HR=0.5, which corresponds to log~N$_H>21.5$, 22.2,
22.7 at z=0, 1, 2, respectively (assuming a power law slope
$\alpha$=0.8).

According to this definition, the fraction of obscured sources to the
total is 0.36.  Of course, since errors on HR are large, especially at
low fluxes, the separation between obscured and unobscured sources has
only a statistical meaning. To evaluate how the fraction of the
obscured sources is affected by the errors of HR, we generated 1000
random source distributions with the same number of sources as the
real data, by assigning to each source a random HR value drawn from a
gaussian distribution with mean equal to the measured HR and $\sigma$
equal the HR error.  We then computed the dispersion of the number of
sources with HR$>0.5$, which turned out to be about 5\%. The dotted
histogram in the right panel of Figure \ref{hrt} is the average of the
simulated (H-S)/(H+S) distributions.  The last point of the simulated
(H-S)/(H+S) distribution deviates from that of the real (H-S)/(H+S)
distribution because the former takes properly into account the upper
limits, which are plotted in the solid histogram at their nominal
values.

\begin{figure}
\begin{center}
\includegraphics[angle=0,height=8truecm,width=8truecm]{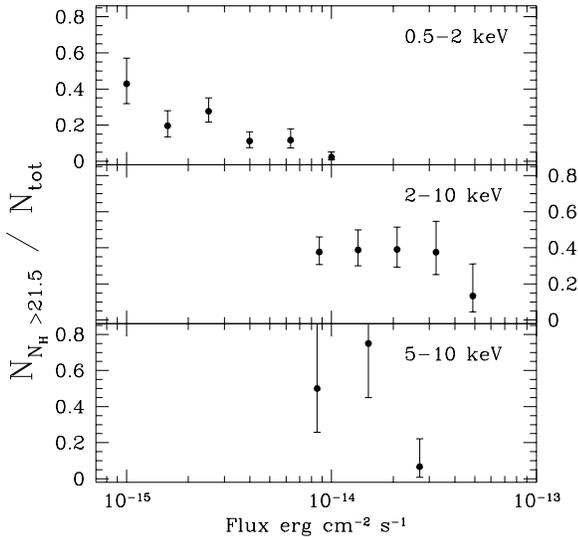}
\caption{Ratio between the number of the obscured
sources to the number of the total sources in the S, H, HH bands (from
top to bottom).}
\label{rabstot}
\end{center}
\end{figure}

Adopting the above definition, Figure \ref{lnlsabuna} shows number
counts of obscured and unobscured sources respectively, while Figure
\ref{rabstot} shows the ratios between the number of the obscured
sources to the total. The fraction of obscured sources increases with
the energy of the X-ray band, consistently with Comastri et
al. (2001). In the 2-10 keV band the fraction of obscured sources
increases from ~$\sim$20\% at fluxes $>(4-5)\times 10^{-14}$ \cgs to
$\sim40\%$ at fluxes~$\sim (1-2) \times 10^{-14}$ \cgs (consistent
with \cite{piconcelli}, \cite{ueda}, \cite{perola},
\cite{lafranca}). It should be beared in mind that all this refers to
moderately obscured sources only, since sources obscured by column
densities N$_H>$a few $10^{24}$ cm$^{-2}$, the so called Compton thick
AGNs, are rarely detected even in 2-10 keV samples (see Tozzi et
al. 2006).

\section {The angular correlation function}

We now focus on the clustering properties of our X-ray
sources. The fraction of active galaxies in our samples is probably
higher than 80-90\% in all bands (according to previous identification
of X-ray sources at similar flux limits, see, e.g., Mainieri et
al. 2002, Fiore et al. 2003, Barger et al. 2005).  We computed the
two point source angular correlation function (ACF, \cite{peebles}) in
the three S, H and F bands using the Landy \& Szalay (LS, 1993)
estimator. The ACF function is given by the following relation:

$$w_{LS}(\theta) = {k_1 \times DD-2 \times k_2\times DR+RR\over RR}, \eqno (1)$$ 

where $DD$ is the number of pairs of sources in the real data with
separation angle in the interval $\theta,\, \theta + d\theta$, $RR$ is
the number of pairs in the same separation angle interval using a
spatially random distribution of sources, $DR$ is the number of pairs
obtained comparing real data and random data, and $k_1 =
{n_r(n_r-1)\over {n_d(n_d-1)}}$, $k_2= {(n_r-1)\over {2 \times n_d}}$,
where n$_d$ is the total number of real sources and n$_r$ is the
number of random sources.

To evaluate $w(\theta)$ we followed two different methods: (1) we
generated 300 random source distributions with the same number of
sources as the real data by convolving a random source distribution
with the differential source number counts and with the appropriate
sensitivity maps.  We adopted a cut in flux for both the real source
and random source distributions to exclude from the ACF analysis the
10\% faintest sources in the samples.  We then computed the ACFs
according to eq. (1) for each random realization and in
logarithmically spaced $\theta$ bins.  The distribution of the 300 ACF
values in each $\theta$ bin is in most cases reasonably well fitted by
a gaussian distribution and therefore we computed the mean and the rms
values in each $\theta$ bin.  (2) we generated a single random source
distribution using 100 times more sources than in the real
samples. Also in this case, we adopted a cut in flux for both the real
source and random source distributions to exclude from the ACF
analysis the 10\% faintest sources and computed the ACFs according to
eq. (1).  In this case errors were computed adopting the following
formula~ (from Landy \& Szalay, 1993):

$$ \delta(\theta) = {(1 + w(\theta)) \over \sqrt{(DD})}  \eqno(2)$$

\begin{figure}
\begin{center}
\includegraphics[angle=0,height=8truecm,width=8truecm]{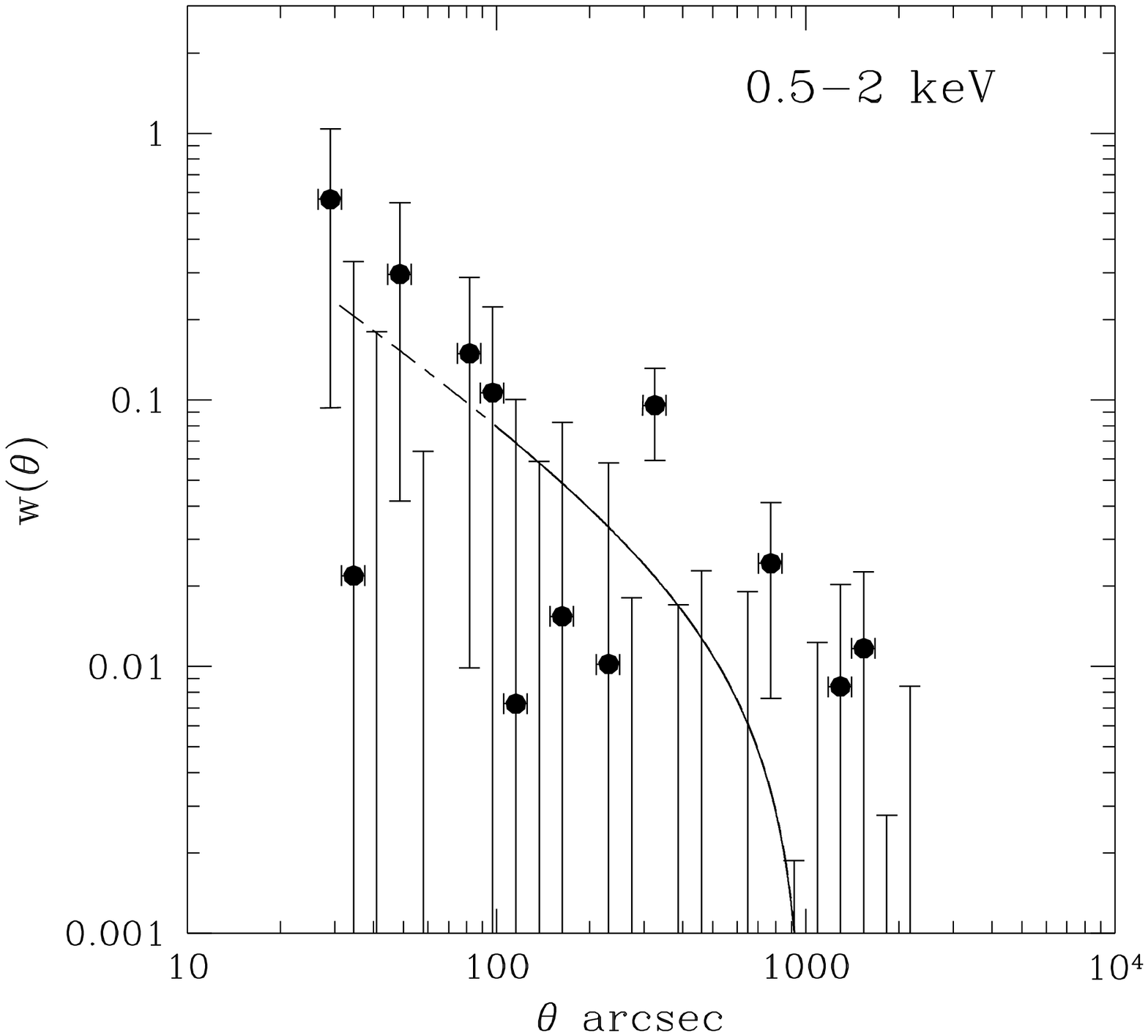}
\includegraphics[angle=0,height=8truecm,width=8truecm]{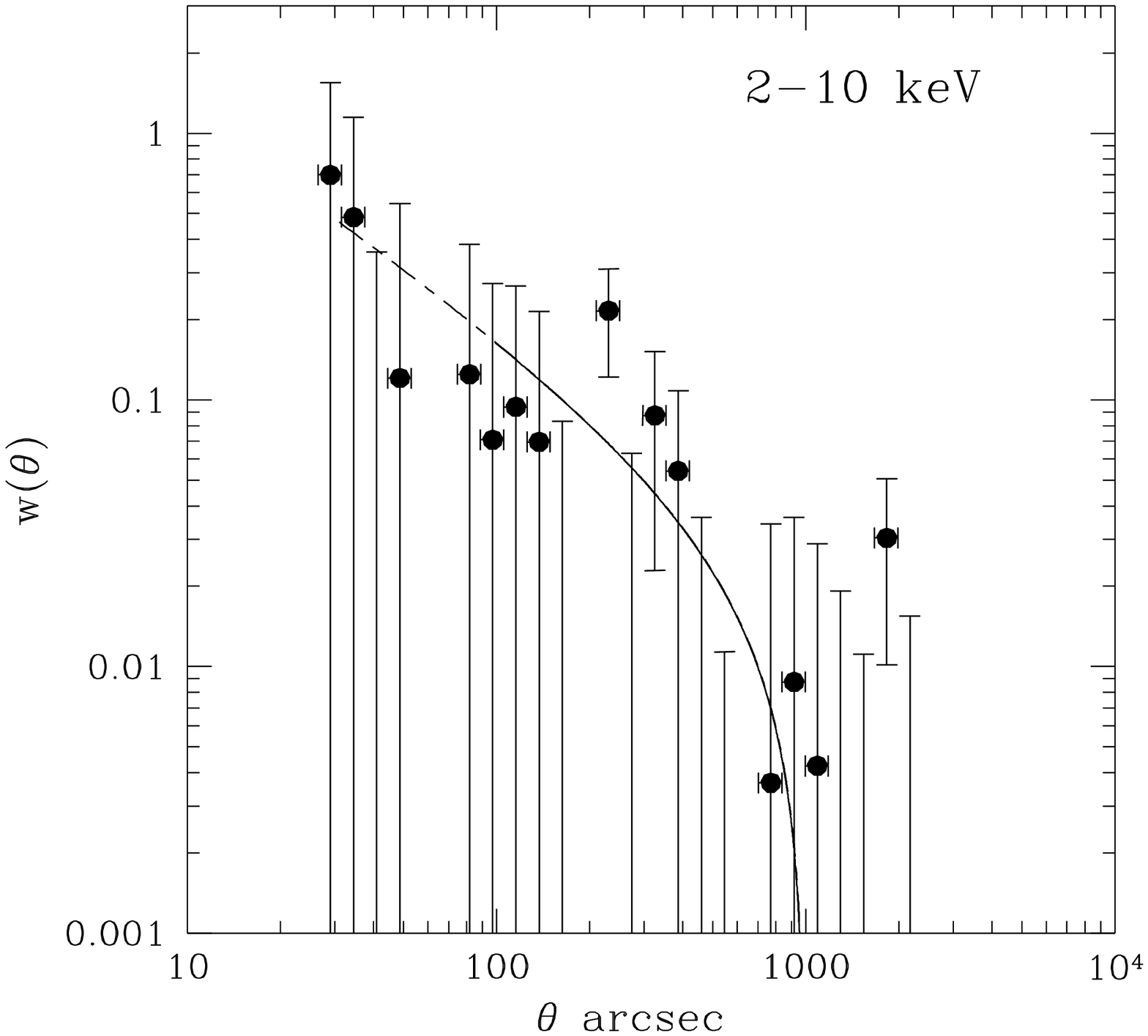}
\caption{The angular correlation function in the S band (top panel),
and H band (bottom panel) computed using the Landy \& Szalay
estimator. The  solid lines represent our best fit to the data between 100
and 1000 arcsec.}
\label{angcorr}
\end{center}
\end{figure}

Figure \ref{angcorr} shows the ACF in the S and H bands computed in
bins of log~$\theta=0.075$ following the second method.  The solid
lines represent the best fit $w(\theta) = (\theta/\theta_0)^{1-\gamma}
- \omega_{\Omega}$ function to the ACF. The $\omega_{\Omega}$ factor
has been introduced following Landy \& Szalay (1993), Daddi et
al. (2000) and Basikalos et al. (2004) to take into account the so
called ``integral constraint''. This results from the fact that the
correlation function is evaluated in a limited area, while the ACF is
normalized to zero over the full sky.  According to \cite{roche} this
implies that the $w(\theta)$ computed using eq. (1) over-estimates the
real $w(\theta)$ by a constant quantity
$\omega_{\Omega}$$=(1/\theta_0)^{1-\gamma}\times C$, where C has been
estimated following again \cite{roche}.  Fixing $\gamma$ to 1.8, as in
\cite{vikh}, Basilakos et al. (2005) and D'Elia et al. (2005),
produces the following best fit correlation angles:
$\theta_0=5.2\pm3.8$ arcsec, $\theta_0=12.8\pm7.8$ and
$\theta_0=7.5\pm3.1$ arcsec in the S, H and F bands respectively
(errors are at 1$\sigma$ confidence level). Letting $\gamma$ to vary
of course increases the already large errors on $\theta_0$ but does
not change their best fit values. In fact the best fit $\gamma$ turn
out to be $1.7^{+0.1}_{-0.5}$ and $1.8^{+0.1}_{-0.2}$ for the S and H
bands respectively. To test the robustness of our determinations we
computed the ACF with different bin sizes from log~$\theta=0.05$ to
log~$\theta=0.1$.  For the 0.5-2 keV ACF we found best fit $\theta_0$
in the range $\sim$4-9 arcsec, while for the 2-10 keV ACF we found
best fit values in the range $\sim$9-19 arcsec.  In all cases the best
fit $\theta_0$ were consistent, within the errors, with the values
given above.

Very similar results are obtained by using the first method to compute the
$w(\theta)$ and the corresponding errors.

Since the correlation angles we probe are of the same order of magnitude of
the XMM-Newton PSF, the so called amplification bias (see, e.g., Vikhlinin \&
Forman 1995) could in principle affect our estimates of w($\theta$) at small
separations, due to sources closer than a few arcsec being observed as a
single source.  However, we note that at our limiting flux the source
confusion affects our images little (see Sect. 2.5). Furthermore, since our
fits have been performed on angular scales $\theta > 100 \arcsec$, much
greater than the XMM-Newton PSF, the amplification bias is not expected to
affect our estimates of the ACF. Indeed, Basilakos et al. (2005) found that
amplification bias is negligible in their XMM-Newton 2dF survey.

\subsection {Source clustering}

The ACF computed in the previous Sect. measures the strength of the
angular correlation, but does not say anything about where the sources
are clustered. To investigate this issue we plot in Figure
\ref{density}a) and \ref{density}b) the source density map of the
0.5-2 keV and 2-10 keV sources, respectively, smoothed with a box of 10
arcmin side. The high density regions of both maps encompass the two
clusters at z$\sim0.4$ (XMMES1\_224 and XMMES1\_374) and other two
clusters or groups of galaxies with a redshift consistent with
0.4. However, these observed source density maps are affected by
instrumental biases, i.e. a higher source density is visible in the
regions with the highest exposure (see Figure \ref{expo}).  To account
for this effect, we computed the deviations of the real source
distribution from a random source one, which has been obtained by
convolving a random source distribution with the differential source
number counts and with the appropriate sensitivity maps. We plot in
Figure \ref{density} c) and \ref{density} d) the normalized source
density $NNSD$ of the 0.5-2 keV and 2-10 keV sources respectively
using the following equation:
\begin{small}
$$NNSD(X,Y)=(NSD(X,Y)-NRD(X,Y))/NRD(X,Y) \eqno(11)$$
\end{small}
where $NSD$ is the real source density and NRD is the density of a
random source sample. The resulting normalized source density map has
then been smoothed with a box of 10 arcmin side.  The high density
regions of the 0.5-2 keV map still encompass the four extended sources
clusters at z$\sim0.4$. On the other hand, the highest density region
of the 2-10 keV map is just located in the central-west region of the
field, in an area which coincides with one of the high density regions
of the 0.5-2 keV map, but in which no extended source has been
found.

\section{Discussion}

We have surveyed with XMM-Newton the central $\sim0.6$ deg$^2$ region of the
ELAIS-S1 field down to flux limits of $\sim 5.5\times 10^{-16}$ \cgs (0.5-2
keV), $\sim 2\times 10^{-15}$ \cgs (2-10 keV) and $\sim 4\times 10^{-15}$ \cgs
(5-10 keV ).  448 sources have been detected in the 0.5-10 keV F band images
with a probability lower than 2$\times$10$^{-5}$ that they are Poisson
fluctuations of the background. 395, 205 and 31 sources are detected in the
0.5-2 keV, 2-10 keV and 5-10 keV band images respectively. Of these, 26 and 4
sources are detected only in the S and H band but not in the F band. The total
number of detected sources is 478.

The source number counts are in good agreement with previous determinations
from large area surveys and compilations (Baldi et al. 2002, Moretti et
al. 2003), showing that field to field variance is smoothed out at the scale
of the ELAIS-S1 survey (0.6 deg$^2$) and implying that the ELAIS-S1 region, at
least from this point of view, can be considered as representative of the
average X-ray source population.  The average spectral properties of the
ELAIS-S1 sources are consistent with those of other surveys covering similar
or larger areas at a similar depth. In particular, we find that: a) the
fraction of obscured sources increases with the energy of the X-ray band and
b) in the 2-10 keV band this fraction increases from ~$\sim$20\% at fluxes
$>(4-5)\times 10^{-14}$ \cgs to $\sim40\%$ at fluxes~$\sim (1-2) \times
10^{-14}$ \cgs (consistent with \cite{piconcelli}, \cite{ueda}, \cite{perola},
\cite{lafranca}).

\begin{figure*}[t!]
\begin{center}
\begin{tabular}{cc}
\includegraphics[angle=0,height=8truecm,width=8.0truecm]{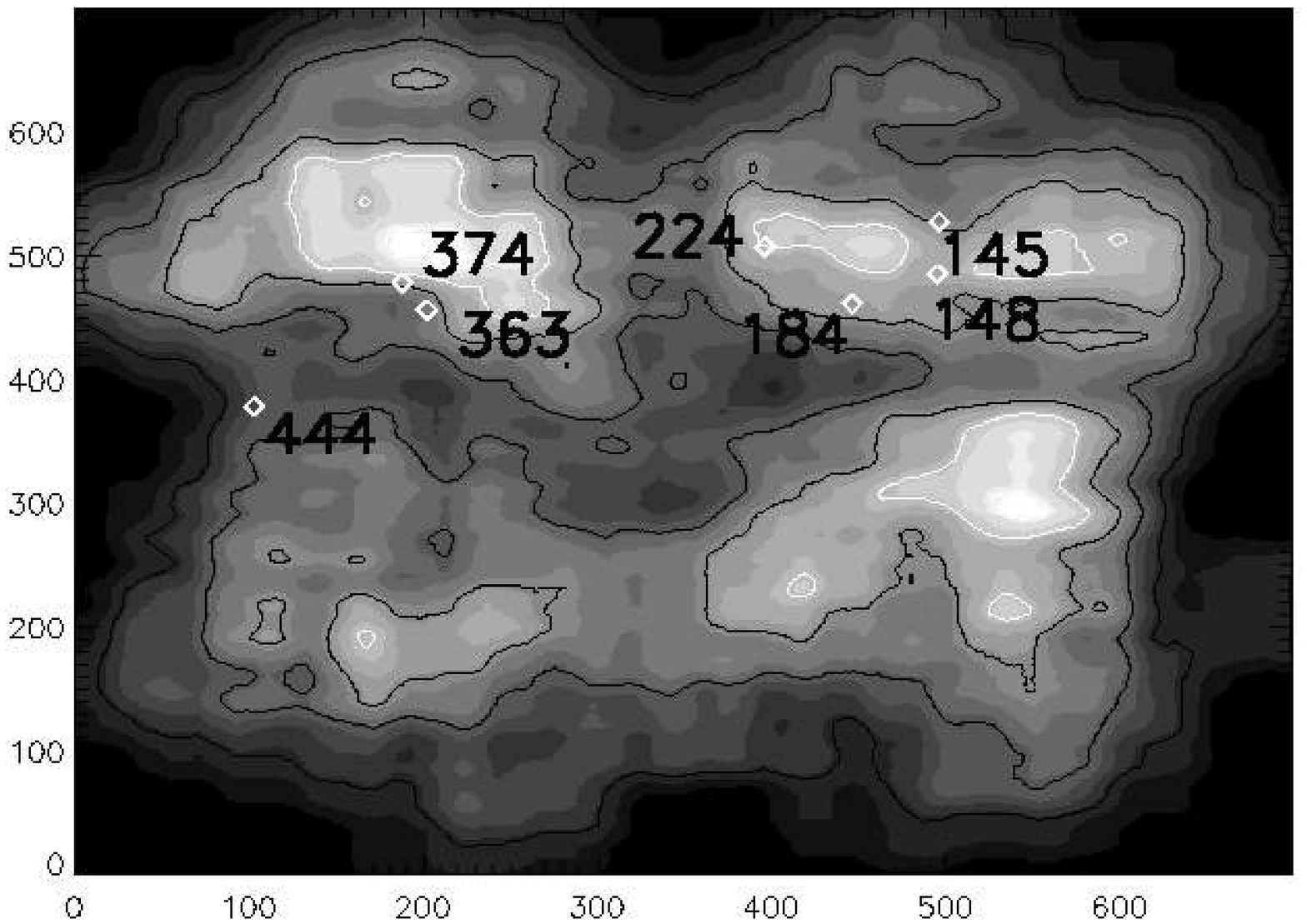}
\includegraphics[angle=0,height=8truecm,width=8.0truecm]{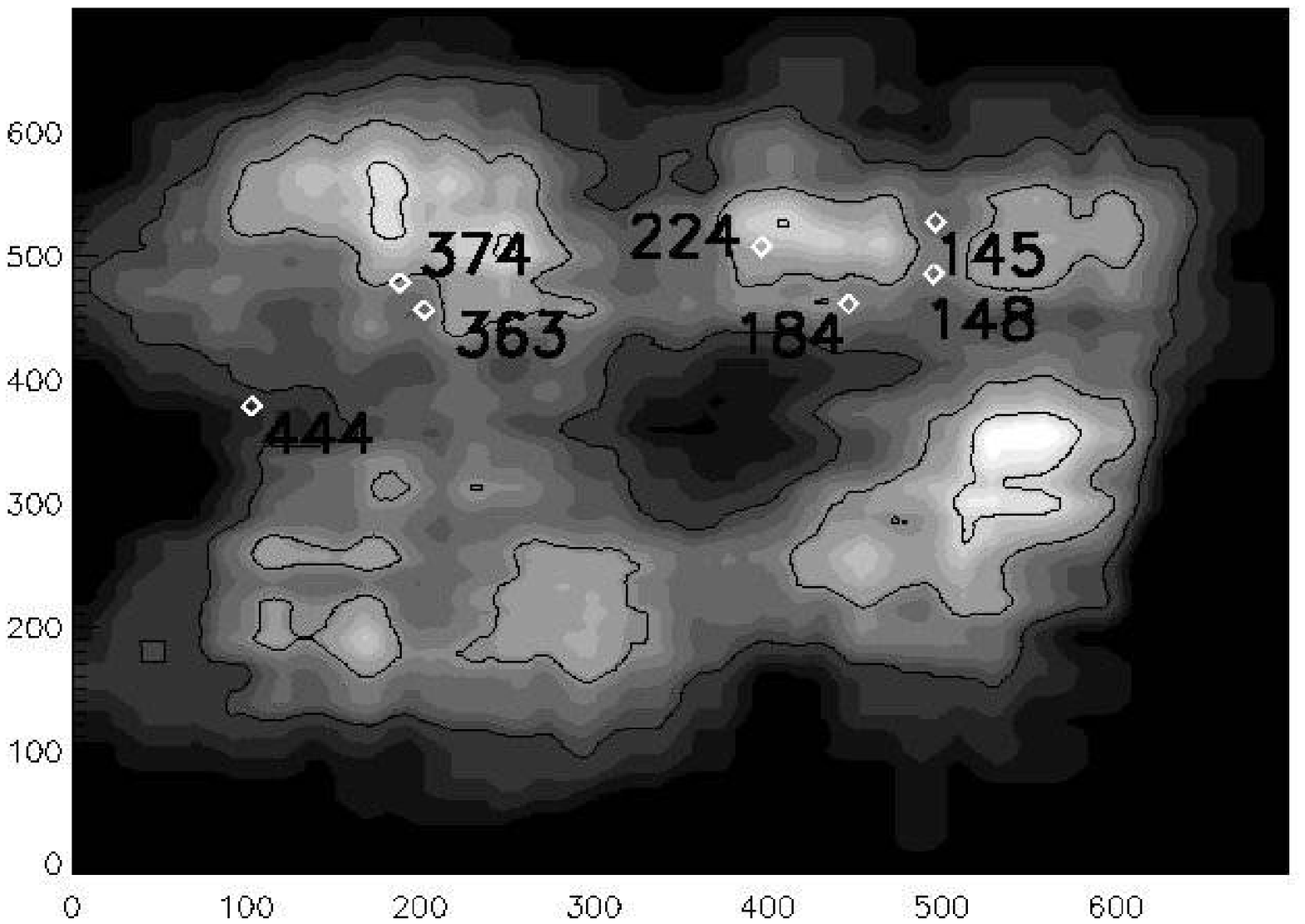}
\end{tabular}
\begin{tabular}{cc}
\includegraphics[angle=0,height=8truecm,width=8.0truecm]{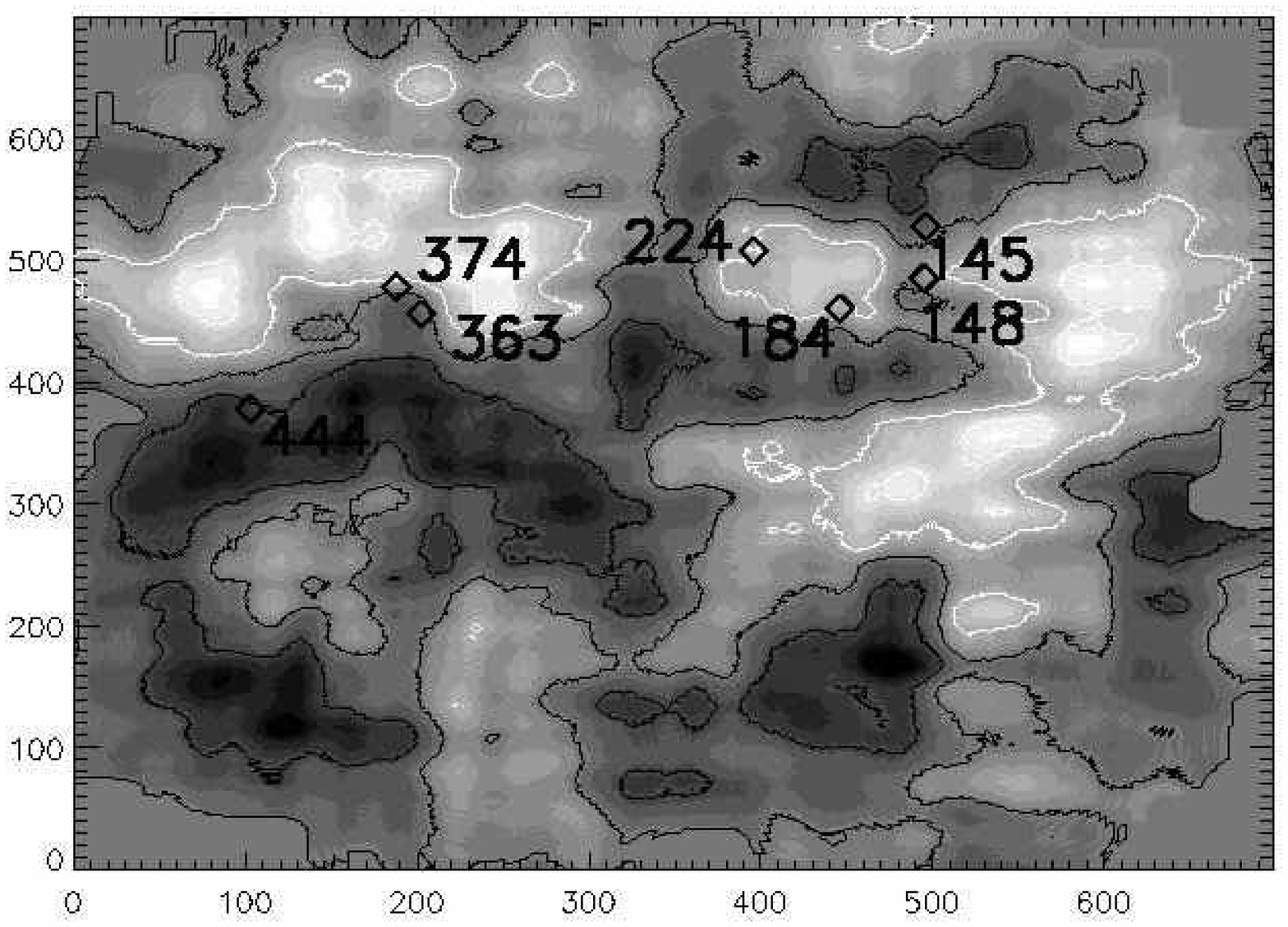}
\includegraphics[angle=0,height=8truecm,width=8.0truecm]{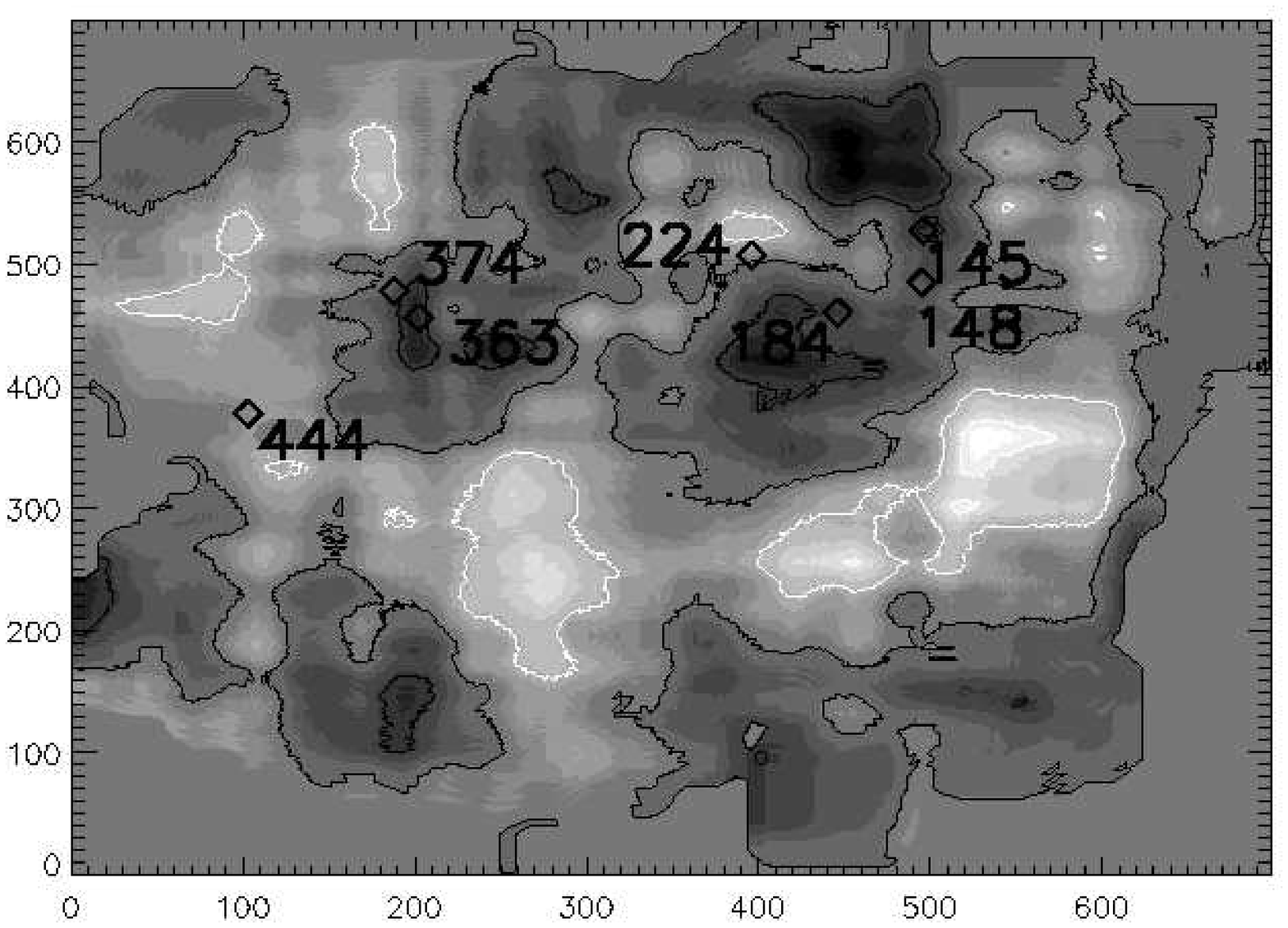}
\end{tabular}
\caption{Top left panel a): the 0.5-2 keV source density
contours, corresponding to -30\%, -15\%, 5\%, 20\% and 40\% of the
average source density; top right panel b): the 2-10 keV
source density contours corresponding to -40\%, -20\%, 5\%, 30\% and
55\% of the average source density. Darker regions correspond to
underdensities and lighter regions to overdensities. The diamonds in
all panels represent the 7 extended sources in the field (see
Sect. 2.5). Bottom left panel c): same as top left panel for the normalized
0.5-2 keV source density; bottom right panel d): same as top right panel for the  2-10 keV  source density contours.
}
\label{density}
\end{center}
\end{figure*}

We identified 7 clearly extended sources in the field and estimated their
redshift through spectral fits with thermal models.  In two cases the
constraint on the redshift is quite accurate, $\sigma (z) /(1+z)<1.5\%$, in
three other cases $\sigma (z) /(1+z)\sim(4-10)\%$, and in the remaining two
cases the upper limit to the redshift is not well determined. Of course these
uncertainties do not take into account possible systematic errors and should
therefore be considered with some caution.  The redshift of XMMES1\_363 is
constrained to be $>0.7$ at the 90\% confidence level.  In four cases
(XMMES1\_224, XMMES1\_374, XMMES1\_184 and XMMES1\_148) the redshift is
consistent with z$\sim$0.4.  Photometric redshifts estimated using the B, V,
R, K and J magnitudes of the brightest galaxies associated with the seven X-ray
sources are all consistent with the redshifts estimated through their X-ray
spectra.  Interestingly, a peak at z$\sim0.4$ is present in the spectroscopic
redshift distribution of radio sources around XMMES1\_374 and XMMES1\_184
(Gruppioni et al. in preparation). The 0.5-10 keV luminosity of XMMES1\_224 is
$\sim8\times 10^{43}$ erg s$^{-1}$, while the luminosity of the other three
extended sources at a similar redshift is 15-60 times smaller.  The angular
separations between the luminous cluster XMMES1\_224 and the 3 lower
luminosity clusters or groups of galaxies are $5'$ (XMMES1\_184), $7.5'$
(XMMES1\_148) and $15.5'$ (XMMES1\_374), corresponding to 1.6, 2.4 and 5 Mpc
at z$\sim$0.4, for the assumed cosmology.  A detailed analysis of the extended
sources and of the large scale structure at z$\sim0.4$, including
spectroscopic redshifts, will be presented in a forthcoming publication.

We have computed the angular correlation function of the sources in
the S and H bands using the Landy \& Szalay (1993) estimator. Fixing the slope
of the correlation function at $\gamma=$1.8, we found
the following correlation angles: $\theta_0=5.2\pm3.8$ arcsec and
$\theta_0=12.8\pm7.8$ arcsec in the S, and H bands respectively, while
the correlation angle in the 0.5-10 keV band is intermediate between
these values. These estimates are fully consistent with the
statistically poorer D'Elia et al. (2005) determinations.  Basilakos
et al. (2005) and Basilakos et al. (2004) found $\theta_0=10.4\pm1.9$
arcsec in the S band and $\theta_0=28\pm9$ in the H band for samples
of sources of size similar to our ELAIS-S1 sample, but with a flux
limit $\sim5-6$ times higher than ours and spread over an area 4 times
wider.

The correlation angle of hard X-ray selected sources is formally
$\sim$2.5 times larger than that of soft X-ray selected sources (a
correlation angle of hard sources higher than soft sources has been
suggested also by Yang et al. 2003 and Basilakos et al 2004), although
the difference is significant at only $\sim1\sigma$ level. If real,
this difference may be due to the different redshift distributions of
hard and soft sources, because the angular correlation angle depends
also on the average distance between galaxies, and therefore on their
redshift. For a proper comparison one would need to evaluate the
spatial correlation function, once the redshifts of the sources are
known. A campaign to obtain optical spectroscopy of the optical
counterparts of the X-ray sources is currently on going using
VIMOS@VLT and the results will reported in a future publication. For
the time being, we can obtain a rough estimate of the present-day
correlation length r$_0$ assuming an appropriate redshift distribution
dN/dz and inverting the Limber equation (\cite{limber},
\cite{peebles}).  We have estimated the observed redshift
distributions of the S and H band sources, by convolving the best fit
luminosity functions of La Franca et al. (2005, 2-10 keV) and Hasinger
et al. (2005, 0.5-2 keV) with ELAIS-S1 sky coverage. Of course this
approach does not take into account the presence of significant
structures at a given redshift in the ELAIS-S1 field which could bias
the redshift distribution of the ELAIS-S1 sources (one such structure
may be present at z$\sim0.4$, see Sect 2.5), and therefore the
following results are should be considered as upper limits. Assuming
an evolutionary parameter $\epsilon$ \footnote{$\epsilon$
parameterizes the redshift evolution of the clustering (see, e.g.,
Overzier et al. 2003).}$=\gamma-3=-1.2$ (for $\gamma=1.8$, the so
called ``comoving clustering model''), which should be appropriate for
active galaxies (\cite{kundic}, \cite{croom}, \cite{croom05}), we
obtain r$_0=12.8^{+3.6}_{-5.7}$ h$^{-1}$ Mpc and
r$_0=17.9^{+4.2}_{-6.1}$ h$^{-1}$ Mpc in the S and H band
respectively. So far we have obtained secure optical
spectroscopy redshifts for 87 and 49 sources in the S and H bands
(Feruglio et al., in preparation).  Using these (admitedly
incomplete) redshift distributions, which clearly shows strong peaks
at z=0.4, we obtain r$_0=9.8^{+2.7}_{-4.3}$ h$^{-1}$ Mpc and
r$_0=13.4^{+3.2}_{-4.6}$ h$^{-1}$ Mpc in the S and H band, respectively.

The S and H band redshift distributions are peaked at z$\sim$1 and
z$\sim$0.9 respectively (median redshift) and therefore we are
effectively measuring the clustering at those redshifts. The
correlation length at a given redshift is: $r_0(z)=r_0 \; (1+z)^{-
{(3+\epsilon-\gamma)\over \gamma}}$, and therefore $r_0(z)=r_0$ for
$\epsilon=-1.2$ and $\gamma=1.8$.


Our correlation lengths are lower but still consistent within the
large errors with those of Basilakos et al. (2005) and Basilakos et
al. (2004). On the other hand, for $\gamma=1.8$ \cite{gilli} find
r$_0\sim$12 h$^{-1}$ Mpc and r$_0\sim$6 h$^{-1}$ Mpc for the faint
sources in the CDFS and CDFN fields respectively. The former
estimate is consistent with our determination.  We also note that
although the median redshift of these sources is similar to the median
redshift assumed above for the ELAIS-S1 sources, their fluxes (and
therefore luminosities) are a factor of 5-10 lower than those of the
sources in the ELAIS-S1 field.  Yang et al. (2006) present a
spatial correlation function analysis of the non-stellar X-ray point
sources in the Chandra Large Area Synoptic X-ray Survey of Lockman
Hole Northwest (CLASXS), covering a total area of 0.4 deg$^2$ down to
a flux limit of 3$\times$10$^{-15}$ \cgs in the 2-8 keV band. They
find r$_0=8.1_{-2.2}^{+1.2}$ h$^{-1}$ Mpc, consistent with our
determination obtained using the observed, but incomplete, redshift
distributions.

We have also compared our correlation lengths to those obtained from
optically selected QSOs, optically and near infrared selected galaxies
and radio sources. Croom et al. (2005) find a low correlation length
in redshift space $s_0=4.8\pm1$ h$^{-1}$ for $\gamma=1.23$.  Grazian
et al. (2004) find a low $r_0=8.6\pm2.0$ h$^{-1}$ Mpc for
$\gamma=1.56$ and for a sample of local (0.02$<$z$<$0.22), optically
selected AGN. Both determination are lower than our values.  On the
other hand, the correlation lengths of the ELAIS-S1 X-ray sources are
comparable with that of Extremely Red Objects (EROs) (see, e.g., Daddi
et al. 2001, r$_0=12\pm3$ h$^{-1}$ Mpc for $\gamma=1.8$ with
1$<$z$<$1.2) and luminous radio sources (see, e.g., Overzier et
al. 2003, r$_0=14\pm3$ h$^{-1}$ Mpc for $\gamma=1.8$ ) at similar
redshifts (z$\sim1$).

We finally investigated whether the source density and clustering (on
scales between 3 and 6 arcmin, where the signal in the ACF is more
significant), are spatially related to the large scale structure
discovered at z$\sim0.4$.  Figure \ref{density} shows that there is a
high normalized source density for the S band sample around the
clusters and groups at z$\sim$0.4 (or with a redshift consistent with
this value) extending both toward East and toward South/West,
suggesting that the structure is complex, with a size comparable to
the full XMM-Newton coverage. The highest normalized source
densities of the H band sources are located in the South West and
central-West regions of the field. The first is nearly coincident with
one of the peaks of the S band source density, while the second has
not a S band counterpart.  The H band density peaks appear to be
unrelated to the position of the extended sources at z$\sim0.4$.

Sensitive observations over a wider area are clearly needed to both
improve the statistics of the correlation functions and to cover {\it
several} structures of the size of 10--20 Mpc in the same area, thus
allowing us to understand whether these are peculiar structures, found
by chance in selected areas of the sky, or whether, rather, they are
representative of the average Large Scale Structure.  The ongoing
COSMOS multiwavelength survey will achieve these goals in the
following years.

\begin{acknowledgements}
Part of this work was supported by MIUR COFIN-03-02-23 and INAF/PRIN
270/2003. We thank Adriano Fontana for allowing us to use his
photometric redshift code.
\end{acknowledgements}

\end{document}